\documentclass[a4paper,12pt]{article}
\pdfoutput=1
\usepackage{amssymb}
\usepackage{amsmath}
\usepackage{amsfonts}
\usepackage{mathtools}
\usepackage{bm}
\usepackage{hyperref,graphicx}
\usepackage{slashed}
\usepackage{color}
\usepackage{indentfirst}
\usepackage{graphicx}
\usepackage{bbm}
\usepackage{csquotes} 
\usepackage{caption}
\usepackage{here}
\usepackage{comment}
\usepackage{listings}
\usepackage[italicdiff]{physics}
\numberwithin{equation}{section} 

\def\ignore#1{{}}

\makeatletter
\newcommand*\rel@kern[1]{\kern#1\dimexpr\macc@kerna}
\newcommand*\widebar[1]{%
  \begingroup
  \def\mathaccent##1##2{%
    \rel@kern{0.8}%
    \overline{\rel@kern{-0.8}\macc@nucleus\rel@kern{0.2}}%
    \rel@kern{-0.2}%
  }%
  \macc@depth\@ne
  \let\math@bgroup\@empty \let\math@egroup\macc@set@skewchar
  \mathsurround\z@ \frozen@everymath{\mathgroup\macc@group\relax}%
  \macc@set@skewchar\relax
  \let\mathaccentV\macc@nested@a
  \macc@nested@a\relax111{#1}%
  \endgroup
}
\makeatother

\renewcommand{\thefootnote}{\arabic{footnote}}
\clearpage
\addtocounter{page}{-1}

\def\ignore#1{{}}

\newcommand{\alp}{\alpha}

\newcommand{\gm}{\gamma}
\newcommand{\Gm}{\Gamma}
\newcommand{\dlt}{\delta}

\newcommand{\ep}{\epsilon}
\newcommand{\tht}{\theta}

\newcommand{\lmd}{\lambda}

\newcommand{\sgm}{\sigma}
\newcommand{\Sgm}{\Sigma}

\newcommand{\omg}{\omega}

\newcommand{\bea}{\begin{eqnarray}}
\newcommand{\eea}{\end{eqnarray}}

\newcommand{\tl}[1]{\tilde{#1}}

\newcommand{\bdm}[1]{{\bm #1}}

\newcommand{\diag}{{\rm diag}}
\newcommand{\der}{\partial}

\newcommand{\id}{\mbox{\boldmath $1$}}

\newcommand{\brkt}[1]{\left( #1 \right)}
\newcommand{\brc}[1]{\left\{ #1 \right\}}
\newcommand{\sbk}[1]{\left[ #1 \right]}

\newcommand{\udl}[1]{\underline{#1}}

\newcommand{\gc}{g_{\rm gc}}
\newcommand{\sRb}{s_{\rm b}}
\newcommand{\sRf}{s_{\rm f}}

\newcommand{\cA}{{\cal A}}

\newcommand{\cD}{{\cal D}}
\newcommand{\cE}{{\cal E}}
\newcommand{\cF}{{\cal F}}

\newcommand{\cN}{{\cal N}}
\newcommand{\cO}{{\cal O}}

\renewcommand{\thefootnote}{\fnsymbol{footnote}}

\begin{document}

\title{
\begin{flushright}
\begin{minipage}{0.25\linewidth}
\normalsize
KEK-TH-2722 \\
KYUSHU-HET-322 \\*[50pt]
\end{minipage}
\end{flushright}
{\Large \bf 

Response of Kaluza-Klein mass spectrum \\
to deformations of rugby-ball compact space
\\*[20pt]}}

\author{
Hajime~Otsuka$^{a}$\footnote{
E-mail address: otsuka.hajime@phys.kyushu-u.ac.jp
}
\ and\
Yutaka~Sakamura$^{b,c}$\footnote{
E-mail address: sakamura@post.kek.jp
}\\*[20pt]
$^a${\it \normalsize
Department of Physics, Kyushu University,}\\ 
{\it \normalsize
744 Motooka, Nishi-ku, Fukuoka, 
819-0395, Japan}\\
$^b${\it \normalsize 
KEK Theory Center, Institute of Particle and Nuclear Studies, KEK,}\\
{\it \normalsize 1-1 Oho, Tsukuba, Ibaraki 305-0801, Japan}\\
$^c${\it \normalsize 
Graduate University for Advanced Studies (Sokendai),}\\
{\it \normalsize 1-1 Oho, Tsukuba, Ibaraki 305-0801, Japan.}
}
\date{}

\maketitle

\centerline{\small \bf Abstract}
\begin{minipage}{0.9\linewidth}
\medskip 
\medskip 
\small

We investigate the response of the Kaluza-Klein (KK) mass spectrum to various deformations of the rugby-ball background  
in 6-dimensional supergravity. 
We derived the mode equations that contain the 3-dimensional scale factor and the lapse function. 
By solving these, we numerically evaluate the KK masses for a bulk scalar and a spinor 
when the background has a nontrivial dependence on the position in the compact space. 
We clarify some qualitative features of the spectrum deformation for some perturbations of the rugby-ball background. 
For perturbations of a supersymmetric background, 
we find that the mass-splitting between bosonic and fermionic modes  
is much smaller than the deviation from the values of the supersymmetric mass eigenvalues. 

\end{minipage}

\renewcommand{\thefootnote}{\arabic{footnote}}
\thispagestyle{empty}
\clearpage

\section{Introduction}
\label{introduction}
Models with large extra dimensions have been attracted some attention 
for various reasons~\cite{Arkani-Hamed:1998jmv,Aghababaie:2003wz,Lust:2019zwm,Montero:2022prj}. 
Some of them predict a small cosmological constant, which is identified with the Casimir energy for the compact space, 
and are discussed in the context of the cosmological constant problem. 
Besides, such large extra dimensions affect the cosmological history at early times. 
If the temperature in the radiation-dominated era~$T_{\rm rad}$ is much higher than the compactification scale~$m_{\rm KK}$, 
the Kaluza-Klein (KK) modes affect the expansion rate of the universe in that era 
because they contribute to the energy density and the pressure of the radiation. 

In our previous papers~\cite{Otsuka:2022rpx,Otsuka:2022vgf,Otsuka:2024xsp}, 
we pursue the time evolution of the space and the background values of the moduli 
in a six-dimensional (6D) model compactified on a sphere~$S^2$ by numerically solving the field equations. 
The model is basically the Salam-Sezgin setup~\cite{Salam:1984cj}, 
which is based on the gauged 6D $\cN=(1,0)$ supergravity (SUGRA)~\cite{Nishino:1984gk,Randjbar-Daemi:1985tdc}. 
We found that if $T_{\rm rad}\gg m_{\rm KK}$, the expansion rate for the three-dimensional (3D) non-compact space deviates 
from the usual four-dimensional (4D) cosmology. 
We also found that the moduli oscillation is induced by the pressure for the two-dimensional (2D) compact space, 
even if the moduli are stabilized at the initial time. 
In these works, we considered a case that there are no branes and the compact space has a spherical symmetry, for simplicity. 
In order to construct a realistic model, however, we need to introduce a 3-brane on which the standard model particles live. \footnote{
Since we consider large extra dimensions, the standard model cannot live in the bulk because the gauge coupling constants are suppressed 
by the large volume of the compact space. 
}

In this paper, we introduce two 3-branes that are set at antipodal points, and analyze the KK mass spectrum. 
They break the spherical symmetry, and the background can have a nontrivial dependence on the position in the compact space. 
This makes the analysis of the KK spectrum complicated. 
In the 6D SUGRA, the number of hypermultiplets~$n_H$ and that of vector multiplets~$n_V$ are constrained 
by the anomaly cancellation condition~$n_H-n_V=244$~\cite{Randjbar-Daemi:1985tdc,Green:1984bx,Kumar:2010ru}. 
This indicates the existence of many hypermultiplets in the bulk. 
As a representative of them, we consider a 6D scalar and a 6D spinor to analyze the KK mass spectrum. 
When the background has the 4D Lorentz symmetry, 
the KK masses are determined solely by the information on the compact space geometry. 
In contrast, for cosmological discussions, the background does not have such a symmetry 
and the scale factor for the 3D non-compact space often 
has a nontrivial dependence on the position of the extra dimensions. 
In such a case, the 3D scale factor also affects the KK masses. 
The information on the KK spectrum is indispensable for calculating the energy density and the pressures of the radiation, 
or the Casimir energy for the compact space. 
In the model we consider, the curvature of the compact space and the magnetic flux threading the extra dimensions also contribute 
to the moduli stabilization, in addition to the scalar potential.  
Such a situation is often encountered in string theory. 
Hence our setup can also be regarded as a simplified version of the string-theoretical models. 

The paper is organized as follows. 
In Sec.~\ref{setup}, we explain our setup and the background field configuration. 
In Sec.~\ref{scalar_case}, the KK mass spectrum for a bulk scalar field is numerically evaluated. 
We derive the mode equations that contain the lapse function and the 3D scale factor. 
In Sec.~\ref{spinor_case}, the KK spectrum for a bulk spinor is considered. 
We also discuss the SUSY-breaking effect on the spectrum 
due to the deviation from the rugby-ball solution. 
Sec.~\ref{summary} is devoted to the summary. 
In Appendix~\ref{flux_qtz}, the flux quantization conditions are summarized. 
In Appendix~\ref{expr:sechsbein}, the limit values of the spin connection at the brane positions are derived. 
In Appendix~\ref{behaviors_near_poles}, we collect the behaviors of the mode functions near the brane positions.

\section{Setup}
\label{setup}
We consider a model based on 6D $\cN=(1,0)$ SUGRA~\cite{Nishino:1984gk,Randjbar-Daemi:1985tdc}. 
The 6D spacetime is compactified on a genus-0 2D manifold, 
which is topologically equivalent to a sphere~$S^2$. 
The 6D coordinates are collectively denoted as $x^M$ ($M=0,1,\cdots,5$) or $(t,x^1,x^2,x^3,y^1,y^2)$, 
where $y^m$ ($m=1,2$) are the coordinates of the 2D compact space. 

The bulk action is given by~\footnote{
Throughout the paper, 
we work in the 6D Planck unit~$M_6=1$, where $M_6$ is the 6D Planck mass. 
} 
\begin{align}
 S_{\rm bulk} &= S_{\rm EH}+S_{\sgm F}+S_{\rm matter}+S_{\rm brane}, 
\end{align}
where $S_{\rm EH}$ is the 6D Einstein-Hilbert action, $S_{\rm \sgm F}$ is the action for the dilaton~$\sigma$ 
and the U(1) gauge field~$A_M$, $S_{\rm matter}$ is the action for the other bulk fields that do not have 
non-vanishing background values, 
and $S_{\rm brane}$ is the brane action localized at $\tht=0,\pi$. 
The explicit forms of $S_{\rm EH}$ and $S_{\sigma F}$ are given by~\footnote{
We follow the notation of Ref.~\cite{Aghababaie:2002be} 
with the redefinition of the dilaton field as $g_{\rm gc}^2e^{-\phi}=e^\sigma$ 
($\phi$ is the dilaton in the notation of Ref~\cite{Aghababaie:2002be}). 
}
\begin{align}
 S_{\rm EH} &= \int d^6x\;\frac{\sqrt{-g^{(6)}}}{2}R^{(6)}, \nonumber\\
 S_{\sgm F} &= \int d^6x\;\sqrt{-g^{(6)}}\brc{-\frac{1}{2}\der^M\sgm\der_M\sgm
 -\frac{e^\sgm}{4g_{\rm gc}^2}F^{MN}F_{MN}-2g_{\rm gc}^4e^{-\sigma}}, 
 \label{6Daction}
\end{align}
where $g^{(6)}$ is the determinant of the 6D metric tensor, 
$R^{(6)}$ is the 6D Ricci scalar, $\sgm$ is a real scalar (dilaton), $F_{MN}\equiv \der_MA_N-\der_NA_M$ is the U(1) field strength, 
and $\gc$ is the gauge coupling constant. 
The brane action~$S_{\rm brane}=\int d^6x\;{\cal L}_{\rm brane}$ is given by~\cite{Burgess:2011mt}
\begin{align}
 {\cal L}_{\rm brane} &= \sqrt{-g^{(4)}}\sbk{\brkt{-\tau_++\xi_+e^\sigma\epsilon^{mn}F_{mn}}\dlt^{(2)}(y-y_+)
 +\brkt{-\tau_-+\xi_-e^\sigma\epsilon^{mn}F_{mn}}\delta^{(2)}(y-y_-)}, 
 \label{cL_brane}
\end{align}
where $g^{(4)}$ is the determinant of the 4D induced metric tensor, 
$\tau_\pm$ and $\xi_\pm$ are the brane tensions and the Fayet-Iliopoulos (FI) parameters 
localized at $y^m=y^m_\pm$ ($m=1,2$), respectively. 
The normalization of the 2D delta function~$\delta^{(2)}(y-y_\pm)$ and the definition of the anti-symmetric tensor~$\epsilon^{mn}$ are given 
in \eqref{def:delta} and \eqref{def:epsilon}, respectively. 
In the following, we focus on the case that $\tau\equiv \tau_+=\tau_-$ and $\xi\equiv \xi_+=\xi_-$, for simplicity. 
The bulk matter action~$S_{\rm matter}$ consists of a scalar part~$S_{\rm scalar}$ and a spinor part~$S_{\rm spinor}$. 
Their explicit forms will be shown later. 
(See \eqref{S_scalar} and \eqref{S_spinor}.)

As the coordinates on the compact space, we choose the spherical ones~$(y^1,y^2)=(\theta,\phi)$, 
where $\tht$ ($0\leq \tht\leq\pi$) and $\phi$ ($0\leq \phi<2\pi$) are the polar and the azimuthal angles, respectively. 
We assume that the non-compact 3D space is isotropic, and the 2D compact space has an axial symmetry. 
Thus, the background depends only on $t$ and $\tht$. 
The background metric is expressed as 
\begin{align}
 ds^2 &= -n^2(t,\tht)dt^2+a^2(t,\tht)d\vec{x}^2+b^2(t,\tht)d\theta^2+c^2(t,\tht)d\phi^2, 
 \label{bkg:metric}
\end{align}
and the background values of the other fields are 
\begin{align}
 \sigma &= \sigma(t,\tht), \nonumber\\
 A_t &=A_t(t,\tht), \;\;\;
 A_i = 0, \;\;\;
 A_\tht = A_\tht(t,\tht), \;\;\;
 A_\phi = A_\phi(t,\tht). 
\end{align}
Thus, the field strength has the following components. 
\begin{align}
 F_{ti} &= 0, \;\;\;
 F_{t\tht} = \dot{A}_\tht-A'_t, \;\;\;
 F_{t\phi} = \dot{A}_\phi, \nonumber\\
 F_{ij} &= F_{i\tht} = F_{i\phi} = 0, \;\;\;
 F_{\tht\phi} = A'_\phi, \label{bkg:F}
\end{align}
where the dot and the prime denote the $t$- and the $\tht$-derivatives, respectively. 

The bulk equation of motion (EOM) for the dilaton~$\sigma$ is 
\begin{align}
 0 &= \frac{1}{\sqrt{-g^{(6)}}}\partial_M\left(\sqrt{-g^{(6)}}\partial^M\sigma\right)
 -\frac{e^\sigma}{4\gc^2} F^{MN}F_{MN} \nonumber\\
 &= -\frac{1}{n^2}\left[\ddot{\sigma}+\left(-\frac{\dot{n}}{n}+\frac{3\dot{a}}{a}+\frac{\dot{b}}{b}+\frac{\dot{c}}{c}\right)\dot{\sigma}\right]
 +\frac{1}{b^2}\left[\sigma''+\left(\frac{n'}{n}+\frac{3a'}{a}-\frac{b'}{b}+\frac{c'}{c}\right)\sigma'\right] \nonumber\\
 &\quad
 +\frac{e^\sigma}{2\gc^2}\left(\frac{F_{t\tht}^2}{n^2b^2}+\frac{\dot{A}_\phi^2}{n^2c^2}-\frac{A_\phi^{\prime 2}}{b^2c^2}\right)
 +2\gc^4e^{-\sigma}, 
\end{align}
and the nontrivial (bulk) Maxwell equations are 
\begin{align}
 0 &= \partial_M\left(\sqrt{-g^{(6)}}e^\sigma F^{Mt}\right) 
 = \der_\tht\left(\frac{a^3c}{nb}e^\sigma F_{t\tht}\right), \nonumber\\
 0 &= \partial_M\left(\sqrt{-g^{(6)}}e^\sigma F^{M\theta}\right) 
 = -\partial_t\left(\frac{a^3c}{nb}e^\sigma F_{t\tht}\right), \nonumber\\
 0 &= \partial_M\left(\sqrt{-g^{(6)}}e^\sigma F^{M\phi}\right) \nonumber\\
 &= -\frac{a^3b}{nc}e^\sigma\left[\ddot{A}_\phi+\left(-\frac{\dot{n}}{n}+\frac{3\dot{a}}{a}+\frac{\dot{b}}{b}-\frac{\dot{c}}{c}+\dot{\sigma}
 \right)\dot{A}_\phi\right] \nonumber\\
 &\quad
 +\frac{na^3}{bc}e^\sigma\left[A_\phi''+\left(\frac{n'}{n}+\frac{3a'}{a}-\frac{b'}{b}-\frac{c'}{c}+\sigma'\right)A'_\phi\right]. 
 \label{Maxwell_eq}
\end{align}
From the first two equations, we find that
\begin{align}
 F_{t\tht} &= \dot{A}_\tht-A'_t = C_{t\tht}\frac{nb}{a^3c}e^{-\sgm}, 
 \label{sol:F_ttht}
\end{align}
where $C_{t\tht}$ is a real constant. 
The last equation in \eqref{Maxwell_eq} determines the time evolution of $A_\phi$.

\section{KK mass spectrum of bulk scalar}
\label{scalar_case}
In this section, we discuss the Kaluza-Klein (KK) mass spectrum of a bulk scalar field~$\Phi$. 
This information is necessary to calculate the Casimir energy or the energy density and the pressure of the radiation in the early universe. 
When we consider the radiation contribution to the time evolution of the universe, 
we assume that the bulk relativistic particles are in the thermal equilibrium achieved by fast motion of the particles
compared to the time scale of the background space expansion. 
In such a case, we can neglect the time-dependence of the background. 
Thus, we treat the background fields~$n$, $a$, $b$, $c$, $\sigma$ and $A_M$ 
as functions of only $\tht$ in the rest of the paper.

The action for a 6D massless complex scalar field~$\Phi$ is 
\begin{align}
 S_{\rm scalar} &= -\int d^6x\;\sqrt{-g^{(6)}}\cD^M\Phi^\dagger \cD_M\Phi+\cdots, 
 \label{S_scalar}
\end{align}
where the ellipsis denotes interaction terms other than the minimal coupling, 
which are irrelevant to the following analysis. 
The covariant derivative~$\cD_M$ is given by
\begin{align}
 \cD_M &\equiv \der_M-i\sRb A_M, 
\end{align}
where $\sRb$ is the R-charge.

\subsection{Background gauge field}
The background values of $A_M$ are obtained by solving \eqref{Maxwell_eq}. 
Since we neglect the time-dependence of the background, we obtain from \eqref{sol:F_ttht} 
\begin{align}
 A_t(\tht) &= -C_{t\tht}\int^\tht d\tht'\;\frac{n(\tht')b(\tht')}{a^3(\tht')c(\tht')}e^{-\sgm(\tht')}. 
 \label{int-expr:A}
\end{align}
To simplify the discussion, we consider a case that $C_{t\tht}=0$, i.e., $A_t=0$, in this paper. 

After dropping the $t$-derivatives, the last equation in \eqref{Maxwell_eq} can be solved as
\begin{align}
 F_{\tht\phi} &= A'_\phi = C_{\tht\phi}\frac{bc}{na^3}e^{-\sgm}, 
 \label{expr:F_thtphi}
\end{align}
where $C_{\tht\phi}$ is an integration constant. 
By integrating this, we obtain the expression of the gauge potential~$A_\phi$. 
The integration constant is determined by the brane FI terms.  
However, we should note that two coordinate patches are necessary to express it over the whole compact space. 
Following the discussion in Appendix~\ref{flux_qtz}, we obtain 
\begin{align}
 A_\phi^{(+)}(\tht) &= C_{\tht\phi}\int_0^\tht d\tht'\;\frac{b(\tht')c(\tht')}{n(\tht')a^3(\tht')}e^{-\sigma(\tht')}-\tl{\xi}, 
 \;\;\;\;\; \brkt{0\leq\tht <\pi} \nonumber\\
 A_\phi^{(-)}(\tht) &= -C_{\tht\phi}\int_\tht^\pi d\tht'\;\frac{b(\tht')c(\tht')}{n(\tht')a^3(\tht')}e^{-\sigma(\tht')}+\tl{\xi}, 
 \;\;\;\;\; \brkt{0<\tht\leq \pi}
 \label{expr:A_phi^pm}
\end{align}
where $\tl{\xi}\equiv \gc^2\xi/\pi$. 
The superscripts~$(\pm)$ label the coordinate patches. 
From the Dirac quantization condition, we have
\begin{align}
 \sRb C_{\tht\phi}\int_0^\pi d\tht\;\frac{b(\tht)c(\tht)}{n(\tht)a^3(\tht)}e^{-\sigma(\tht)}-2\sRb\tl{\xi}
 &= k_{\rm b}, 
 \label{Dirac_qtm:boson2}
\end{align}
where $k_{\rm b}\in{\mathbb Z}$. 
(See \eqref{Dirac_qtm:boson}.)
This condition determines the constant~$C_{\tht\phi}$.

\subsection{Mode equation}
The (bulk) EOM for $\Phi$ is 
\begin{align}
 0 &= \frac{1}{\sqrt{-g^{(6)}}}\der_M\brc{\sqrt{-g^{(6)}} g^{MN}\brkt{\der_N\Phi-i\sRb A_N\Phi}}
 -i\sRb A^M\brkt{\der_M\Phi-i\sRb A_M\Phi} \nonumber\\
 &= \brkt{-\frac{1}{n^2}\der_t^2+\frac{1}{a^2}\nabla^2}\Phi
 +\frac{1}{b^2}\sbk{\der_\tht^2+\brkt{\frac{n'}{n}+\frac{3a'}{a}-\frac{b'}{b}+\frac{c'}{c}}\der_\tht}\Phi \nonumber\\
 &\quad
 +\frac{1}{c^2}\sbk{\der_\phi^2-2i\sRb A_\phi\der_\phi-(\sRb A_\phi)^2}\Phi, 
\end{align}
where 
\begin{align}
 \nabla^2 &\equiv \sum_{i=1}^3\frac{\der^2}{\der x^{i\,2}}. 
\end{align}

Now we expand the 6D field~$\Phi$ into the KK modes. 
Since $\Phi$ is periodic for $\phi$ with the periodicity~$2\pi$, we can expand it as
\begin{align}
 \Phi(x^\mu,\tht,\phi) &= \sum_{p,q} \frac{e^{iq\phi}}{\sqrt{2\pi}} f_p(\tht;q)\phi_{p,q}(x^\mu), 
 \label{KKexpand:boson}
\end{align}
where $x^\mu$ is the 4D coordinate, and $p$ and $q$ are integers that label the KK modes. 
Then the above EOM becomes
\begin{align}
 \sum_{p,q} \frac{e^{iq\phi}}{\sqrt{2\pi}} \sbk{f_{p,q}(\tht,\phi)\brkt{-\frac{1}{n^2}\der_t^2\phi_{p,q}
 +\frac{1}{a^2}\nabla^2\phi_{p,q}}+\brkt{\cO_q f_p}\phi_{p,q}} &= 0, 
 \label{lin_EOM}
\end{align}
where the differential operator~$\cO_q$ is defined as 
\begin{align}
 \cO_q &\equiv \frac{1}{b^2}\sbk{\der_\tht^2+\brkt{\frac{n'}{n}+\frac{3a'}{a}-\frac{b'}{b}+\frac{c'}{c}}\der_\tht
 -\frac{b^2}{c^2}\brkt{q-\sRb A_\phi}^2}. 
\end{align}
We choose the mode functions~$f_p(\tht;q)$ as solutions of 
\begin{align}
 \cO_q f_p(\tht;q) &= -\frac{m_{p,q}^2}{n^2(\tht)}f_p(\tht;q), 
 \label{md_eq:f}
\end{align}
where $m_{p,q}$ are the KK masses. 
Since all the coefficients in $\cO_q$ are real, we can choose $f_p(\tht;q)$ as a real function without the loss of generality. 
Using \eqref{md_eq:f}, the EOM~\eqref{lin_EOM} becomes
\begin{align}
 \sum_{p,q} \frac{e^{iq\phi}}{\sqrt{2\pi}} f_p(\tht;q)\brkt{-\frac{1}{n^2(\tht)}\der_t^2+\frac{1}{a^2(\tht)}\nabla^2-\frac{m_{p,q}^2}{n^2(\tht)}}\phi_{p,q}(x^\mu) &= 0. 
 \label{EOM:KKexpand}
\end{align}
Here we rescale the coordinates~$x^i$ ($i=1,2,3$) and $\tht$ as
\begin{align}
 \tilde{x}^i &\equiv \frac{a(\tht)}{n(\tht)}x^i, \;\;\;\;\;
 \tilde{\tht} \equiv \tht, 
 \label{rescaling}
\end{align}
which leads to~\footnote{
This coordinate redefinition can cause an additional contribution to the coefficient of $\der_{\tl{x}^i}^2$ through 
$\der_{\tht}=\der_{\tl{\tht}}+(a/n)'x^i\der_{\tl{x}^i}$. 
However, such a contribution does not appear because \eqref{EOM:KKexpand} does not have $\der_\tht$-dependent terms anymore. 
We have already eliminated it by using \eqref{md_eq:f}. 
}  
\begin{align}
 \frac{\der}{\der x^i} &= \frac{a(\tht)}{n(\tht)}\frac{\der}{\der\tl{x}^i}. 
\end{align}
Then \eqref{EOM:KKexpand} is rewritten as
\begin{align}
 \sum_{p,q} \frac{e^{iq\phi}}{\sqrt{2\pi}} \frac{f_p(\tht;q)}{n^2(\tht)}\brkt{-\der_t^2+\tilde{\nabla}^2-m_{p,q}^2}\phi_{p,q}(t,\tl{x}^i) =0, 
\end{align}
where 
\begin{align}
 \tilde{\nabla}^2 &\equiv \sum_{i=1}^3\frac{\der^2}{\der\tilde{x}^{i\,2}}. 
\end{align}
Using the orthonormal relation,~\footnote{
The weight function~$a^3bc/n$ is determined by the coefficient of $\der_\tht$ in $\cO$. 
} 
\begin{align}
 &\quad
 \int d\tht d\phi\;\frac{a^3bc}{n}\brc{\frac{e^{iq\phi}}{\sqrt{2\pi}}f_p(\tht;q)}^*\brc{\frac{e^{iq'\phi}}{\sqrt{2\pi}}f_{p'}(\tht;q')} \nonumber\\
 &= \dlt_{q,q'}\int_0^\pi d\tht\;\frac{a^3(\tht)b(\tht)c(\tht)}{n(\tht)}f_p(\tht;q)f_{p'}(\tht;q') = \dlt_{q,q'}\delta_{p,p'}, 
 \label{orthonormal:f}
\end{align}
we obtain the 4D EOM, 
\begin{align}
 \brkt{\der_t^2-\tilde{\nabla}^2+m_n^2}\phi_n &= 0. 
\end{align}
In terms of the original coordinates, this becomes 
\begin{align}
 \brkt{\der_t^2-\frac{n^2(\tht)}{a^2(\tht)}\nabla^2+m_n^2}\phi_n = 0. 
\end{align}
Therefore, the energy of a 4D particle with the 3D momentum~$\vec{P}$ and the KK level~$(p,q)$ is given by
\begin{align}
 \cE_{\vec{P},p,q} &= \sqrt{\frac{n^2(\tht)}{a^2(\tht)}\sum_{i=1}^3 P_i^2+m_{p,q}^2}. 
 \label{dispersion_rel}
\end{align}

\subsection{Boundary conditions and KK mass spectrum} \label{BC_and_KKmass}
The KK masses~$m_{p,q}$ are determined by the boundary conditions at $\tht=0,\pi$. 
Here we should note that the KK expansion is performed separately in each patch, 
and thus \eqref{md_eq:f} has the form of 
\begin{align}
 \sbk{\der_\tht^2+\brkt{\frac{n'}{n}+\frac{3a'}{a}-\frac{b'}{b}+\frac{c'}{c}}\der_\tht-\frac{b^2\brkt{q^{(\pm)}-\sRb A_\phi^{(\pm)}}^2}{c^2}+\frac{b^2\lmd_q^2}{n^2}}
 f^{(\pm)}(\tht;q^{(\pm)}) &= 0,  
 \label{md_eq:f^pm}
\end{align}
where $\lmd_q$ is a real constant. 
Since the differential operator~$\cO_q$ in \eqref{md_eq:f} is independent of the KK label~$p$, we have dropped it. 
It labels the infinite number of solutions of \eqref{md_eq:f^pm}. 
From \eqref{transition}, \eqref{Dirac_qtm:boson} and \eqref{KKexpand:boson}, the KK labels~$q^{(\pm)}$ are related as
\begin{align}
 q^{(+)} &= q^{(-)}+k_{\rm b}. 
\end{align}
From \eqref{dervartht} and \eqref{Dirac_qtm:boson}, we have
\begin{align}
 \sRb\brkt{A_\phi^{(+)}-A_\phi^{(-)}} &= k_{\rm b}. 
\end{align}
Thus, it follows that~\footnote{
This quantity corresponds to $-i\cD_\phi$. 
}
\begin{align}
 q^{(+)}-\sRb A^{(+)}_\phi &= q^{(-)}-\sRb A^{(-)}_\phi. 
\end{align}
Therefore, the mode equation has the same form in both patches, and is written as
\begin{align}
 \sbk{\der_\tht^2+\brkt{\frac{n'}{n}+\frac{3a'}{a}-\frac{b'}{b}+\frac{c'}{c}}\der_\tht-\frac{b^2\brkt{q-\sRb A^{(+)}_\phi}^2}{c^2}+\frac{b^2\lmd_q^2}{n^2}}
 f(\tht;q) &= 0, 
 \label{md_eq:g:q0}
\end{align}
where $q=q^{(+)}$. 
We can numerically integrate this 
with the proper boundary conditions for $f(\tht=0;q)$ and $f'(\tht=0;q)$, and obtain $f(\tht=\pi;q)$, which implicitly depends on $\lmd_q$. 
Only when $\lmd_q$ equals to a KK mass eigenvalue, 
the integrated value~$f(\pi;q)$ satisfies the boundary condition at $\tht=\pi$. 
Since the normalization of $f(\tht;q)$ does not affect the determination of the KK masses, 
we can freely choose the overall normalization of the ``initial boundary conditions'' at $\tht=0$. 
From the discussions in Appendix~\ref{behaviors:scalar}, 
the initial conditions at $\tht=\ep$ ($0<\ep\ll 1$) can be chosen as~\footnote{
Note that $f'(0;q)$ diverges when $\abs{\zeta_{\rm b}}<1$. 
Hence we impose the initial conditions at $\tht=\ep$ rather than $\tht=0$. 
}
\begin{align}
 f(\ep;q) &= \ep^{\abs{\zeta_{\rm b}}}, \;\;\;\;\;
 f'(\ep;q) = \abs{\zeta_{\rm b}}\ep^{\abs{\zeta_{\rm b}}-1}, 
\end{align}
where 
\begin{align}
 \zeta_{\rm b} &\equiv \frac{q+\sRb\tl{\xi}}{r_0}, \;\;\;\;\;
 r_0 \equiv \lim_{\tht\to 0}\frac{c(\tht)}{\tht b(\tht)}, 
\end{align}
and if we define 
\begin{align}
 \cF_q(\lmd_q) &\equiv \lim_{\tht\to\pi}\brkt{\pi-\tht}^{\abs{\alp_{\rm b}}}f(\tht;q), 
 \label{def:cF}
\end{align}
where 
\begin{align}
 \alp_{\rm b} &\equiv \frac{q-k_{\rm b}-s_{\rm b}\tl{\xi}}{r_\pi}, \;\;\;\;\;
 r_\pi \equiv \lim_{\tht\to\pi}\frac{c(\tht)}{(\pi-\tht)b(\tht)}, 
\end{align}
this is regular at $\tht=\pi$. 
Then, the KK eigenvalues~$\lmd_q=m_{p,q}$ are obtained as solutions of $\cF_q(\lmd_q)=0$.\footnote{
Note that $\cF_q(\lmd_q)$ is an oscillating function.  
Thus the equation~$\cF_q(\lmd_q)=0$ has infinitely many solutions labeled by the integer~$p$. 
}

\subsection{Explicit examples}
In order to illustrate the effects of the background metric on the spectrum, 
we numerically evaluate the KK masses for some specific backgrounds. 

\subsubsection{Static spherical case}
In the absence of the brane (i.e., $\tau=\xi=0$), the following spherical symmetric solution exists~\cite{Salam:1984cj}. 
\begin{align}
 n(\tht) &= 1, \;\;\;
 a(\tht) = a_{\rm c}, \;\;\;
 b(\tht) = b_{\rm c}, \;\;\;
 c(\tht) = b_{\rm c}\sin\tht, \;\;\;
 \sigma(\tht) = \sigma_{\rm c}. 
 \label{spherical_sol}
\end{align}
where $a_{\rm c}$, $b_{\rm c}$ and $\sigma_{\rm c}$ are constants. 
For a U(1) neutral scalar ($\sRb=0$), the mode equation~\eqref{md_eq:g:q0} becomes
\begin{align}
 \brkt{\der_\tht^2+\cot\tht\der_\tht-\frac{q^2}{\sin^2\tht}+b_{\rm c}^2\lmd_q^2}f(\tht;q) &= 0. 
\end{align}
In this case, it is well known that the KK mass eigenvalues are 
\begin{align}
 \lmd_q &= m_{p,q} = \frac{\sqrt{(\abs{q}+p)(\abs{q}+p+1)}}{b_{\rm c}}, 
 \label{unwarped_case}
\end{align}
where $p=0,1,2,\cdots$. 
If we redefine the KK label as $l\equiv \abs{q}+p$, this becomes the familiar form, 
\begin{align}
 m_{l,q} &= \frac{\sqrt{l(l+1)}}{b_{\rm c}}, 
 \label{unwarped_case2}
\end{align}
where $l=0,1,2,\cdots$ and $-l\leq q\leq l$. 
Thus, there are $(2l+1)$ degenerate states for each $l$.

\subsubsection{Effect of nontrivial $\tht$-dependence of background} \label{warping:scalar}
Here we consider a case that the background has a nontrivial $\tht$-dependence. 
In order to see this effect, 
we add small $\tht$-dependent terms to the spherically symmetric solution~\eqref{spherical_sol}. 

\begin{figure}[t]
\begin{center}
 \includegraphics[scale=0.55]{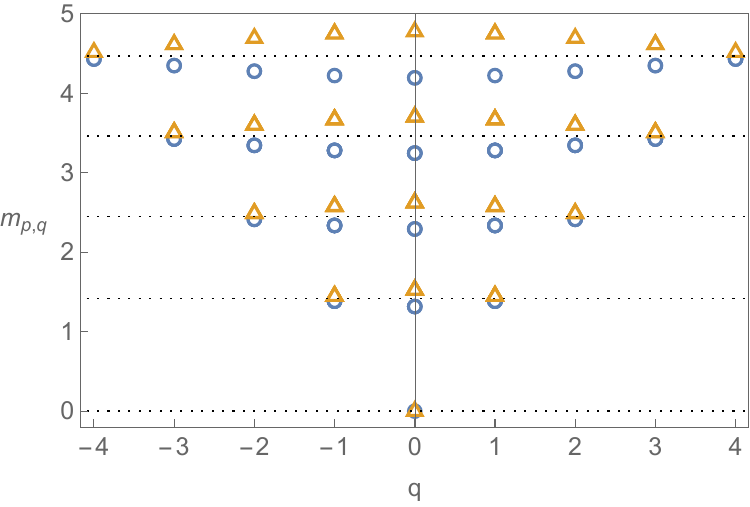} \;\;\;
 \includegraphics[scale=0.55]{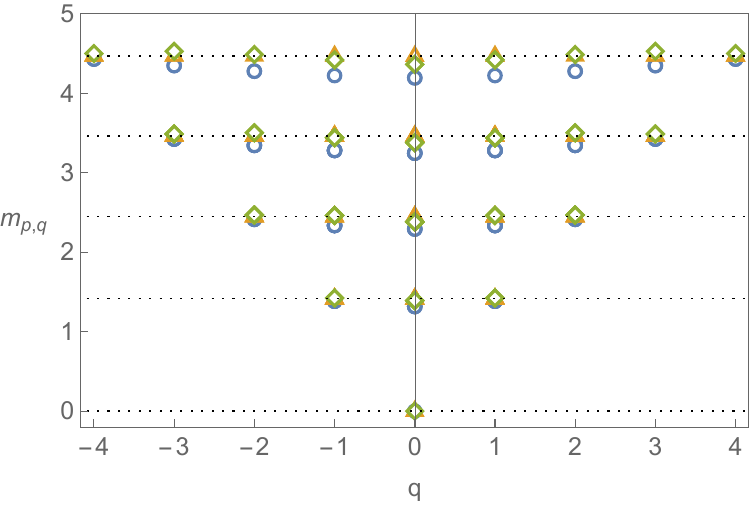}
\end{center}
 \caption{The KK mass eigenvalues~$m_{p,q}$ in the case of \eqref{devi_from_sph} with $b_{\rm c}=1$. 
 The left plot shows the case of $\dlt_b=\pm 0.1$ and $K_b=1$. 
 The (blue) circles and the (orange) triangles correspond to the case of $\dlt_b=0.1$ and $\dlt_b=-0.1$, respectively. 
 The right plot shows the case of $\dlt_b=0.1$ and $K_b=1,2,3$. 
 The (blue) circles, the (orange) triangles and the (green) diamonds correspond to the case of $K_b=1$, $K_b=2$ and $K_b=3$, respectively. 
 The horizontal dotted lines denote the KK masses in the spherical unwarped case~\eqref{unwarped_case2}. 
 }
\label{fig:warped-b1}
\end{figure}

\begin{figure}[t]
\begin{center}
 \includegraphics[scale=0.6]{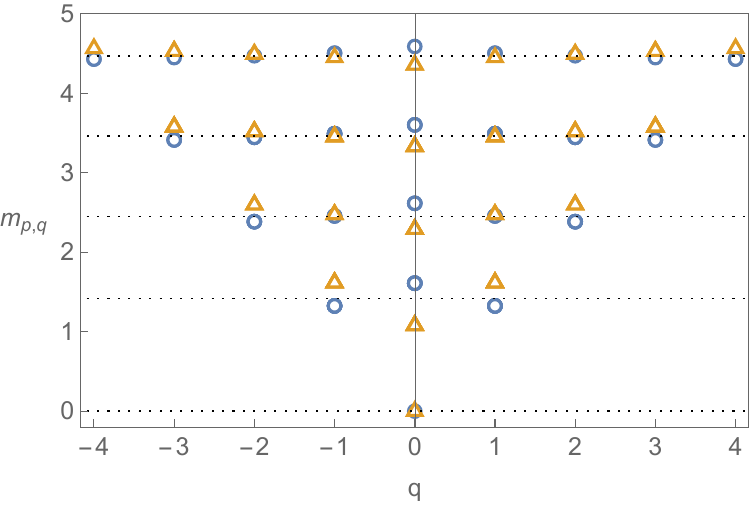}
\end{center}
 \caption{The KK mass eigenvalues~$m_{p,q}$ in the case of \eqref{devi_from_sph:2} with $b_{\rm c}=1$. 
 The (blue) circles and the (orange) triangles correspond to the case of $\dlt_a=0.4$ and $\dlt_a=-0.4$, respectively. 
 The horizontal dotted lines denote the KK masses in the spherical unwarped case~\eqref{unwarped_case2}. 
 }
\label{fig:warped-a04}
\end{figure}

First we consider a case that $b(\tht)$ in \eqref{spherical_sol} is replaced with
\begin{align}
 b(\tht) &= b_{\rm c}\sbk{1+\dlt_b\sin(K_b\tht)}, 
 \label{devi_from_sph}
\end{align}
where $\dlt_b$ is a real constant and $K_b$ is an integer.  
The other background components are unchanged. 
Roughly speaking, the KK mass scale is determined by the volume of the compact space. 
Hence, the deformation of the metric component $b$ or $c$ directly affects the KK mass spectrum. 
Fig.~\ref{fig:warped-b1} shows the KK masses in this case. 
We can see that the degeneracy is resolved due to the violation of the spherical symmetry. 
When $\dlt_b>0$ ($\dlt_b<0$), the volume of the compact space is larger (smaller) compared to the spherically symmetric case, 
and thus the KK masses become lighter (heavier) than \eqref{unwarped_case2} (see the left plot). 
We can also see that the deviation from the degenerate values~\eqref{unwarped_case2} 
becomes smaller for lower KK excitation modes, 
and the $q=0$ states feel a larger effect from the $\tht$-dependence of $b$. 
The right plot in Fig.~\ref{fig:warped-b1} shows the case of $\dlt_b=0.1$, and $K_b=1,2,3$. 
When $K_b\geq 2$, $b(\tht)$ deviates from $b_{\rm c}$ positively or negatively from place to place, 
and the effect of the deformation is averaged out to small. 
As a result, the mass deviation is suppressed compared to the $K_b=1$ case, 

Next we consider a case that $a(\tht)$ in \eqref{spherical_sol} is replaced with 
\begin{align}
 a(\tht) &= a_{\rm c}\brkt{1+\dlt_a\sin\tht}, 
 \label{devi_from_sph:2}
\end{align}
where $\dlt_a$ is a real constant. 
The other background components are unchanged. 
In contrast to the deformation of $b$, that of the 3D scale factor~$a$ does not change the volume of the compact space. 
Nevertheless, it can also change the KK mass spectrum through the coefficient of $\der_\tht$ in \eqref{md_eq:g:q0} 
when $a$ has a nontrivial $\tht$-dependence. 
Fig.~\ref{fig:warped-a04} shows the KK masses in the case of $\dlt_a=\pm 0.4$.\footnote{
The effect of the nontrivial $\tht$-dependence of $a$ is smaller than that of $b$. 
So we take a larger value of $\dlt_a$ in order to magnify the effect. 
}  
We can see that the lower KK modes receive a larger effect of the deformation, 
in contrast to the case of the $b$-deformation. 

\subsubsection{Rugby-ball background}
In the presence of the branes, there are conical singularities at the brane positions. 
Here we consider the so-called rugby-ball solution~\cite{Carroll:2003db,Aghababaie:2003wz,Burgess:2004ib}, 
for which the background is given by 
\begin{align}
 n(\tht) &= 1, \;\;\;
 a(\tht) = a_{\rm c}, \;\;\;
 b(\tht) = b_{\rm c}, \;\;\;
 c(\tht) = r_{\rm c}b_{\rm c}\sin\tht, \;\;\;
 \sigma(\tht) = \sigma_{\rm c}, 
 \label{bkg:simple_case}
\end{align}
where $a_{\rm c}$, $b_{\rm c}$ and $r_{\rm c}$ are positive constants. 
In the presence of the branes at the poles ($\tht=0,\pi$), $r_{\rm c}\neq 1$ and the mass spectrum deviates from \eqref{unwarped_case}. 
The constant~$r_{\rm c}$ is related to the deficit angle~$\dlt_{\rm df}$ as
\begin{align}
 \dlt_{\rm df} &= 2\pi\brkt{1-r_{\rm c}}. 
 \label{rel:dlt-rc}
\end{align}

\begin{figure}[t]
\begin{center}
 \includegraphics[scale=0.6]{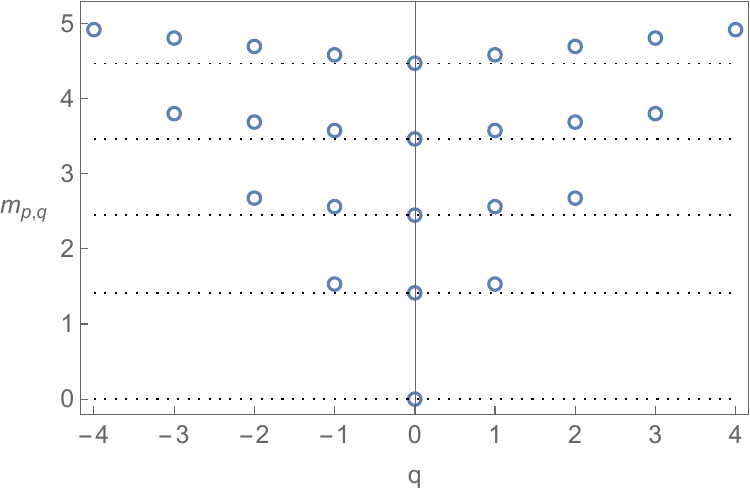}
\end{center}
 \caption{The KK mass eigenvalues~\eqref{expr:barm} in the case of $b_{\rm c}=1$ and $r_{\rm c}=0.9$. 
 The horizontal dotted lines denote the KK masses in the spherical unwarped case~\eqref{unwarped_case2}, i.e., $r_{\rm c}=1$. 
 }
\label{fig:mpq-zeta:0}
\end{figure}

For a U(1) neutral scalar (i.e., $\sRb=0$), the mode equation~\eqref{md_eq:g:q0} is reduced to 
\begin{align}
 \brkt{\der_\tht^2+\cot\tht\der_\tht-\frac{q^2}{r_{\rm c}^2\sin^2\tht}+b_{\rm c}^2\lmd_q^2}f(\tht;q) &= 0, 
\end{align}
whose solutions are expressed in terms of the associated Legendre functions~$P_\mu^\alp(z)$ and $Q_\mu^\alp(z)$. 
Requiring that the mode functions are finite at the poles, we find that they are analytically expressed as
\begin{align}
 f_p(\tht;q) 
 &= \sqrt{\frac{p!(2q/r_{\rm c}+2p+1)}{2\Gm(2q/r_{\rm c}+p+1)}}
 \sbk{\brkt{\cos\frac{\pi q}{r_{\rm c}}}P_{q/r_{\rm c}+p}^{q/r_{\rm c}}(\cos\tht)-\frac{2}{\pi}\brkt{\sin\frac{\pi q}{r_{\rm c}}}Q_{q/r_{\rm c}+p}^{q/r_{\rm c}}(\cos\tht)}, 
\end{align}
where $\Gm(z)$ is the Euler's gamma function. 
This satisfies the orthonormal condition, 
\begin{align}
 \int_0^\pi d\tht\;\sin\tht f_p(\tht;q)f_{p'}(\tht;q) &= \dlt_{p,p'}. 
\end{align}
The corresponding KK masses are given by
\begin{align}
 m_{p,q} &= \frac{1}{b_{\rm c}}\sqrt{\brkt{\frac{|q|}{r_{\rm c}}+p}\brkt{\frac{|q|}{r_{\rm c}}+p+1}}, 
 \label{expr:barm}
\end{align}
where $p=0,1,2,\cdots$. 
Fig.~\ref{fig:mpq-zeta:0} shows the KK mass eigenvalues in the case of $b_{\rm c}=1$ and $r_{\rm c}=0.9$. 
In contrast to Figs.~\ref{fig:warped-b1} and \ref{fig:warped-a04}, the deviation from the spherical limit~\eqref{unwarped_case} 
increases for larger values of $q$. 
When $q=0$, the KK mass becomes equal to that of the spherical case~$m_{p,0}=\sqrt{p(p+1)}/b_{\rm c}$.

In the absence of the branes, (i.e., $r=1$), the mode function~$e^{iq\phi}f_p(\tht;q)/\sqrt{2\pi}$  
reduces to the spherical harmonics~$Y_l^q(\tht,\phi)$ if we identify 
\begin{align}
 l &= \frac{|q|}{r}+p = |q|+p, 
\end{align} 
which satisfy $l=0,1,2,\cdots$ and $-l\leq q\leq l$.

Next we consider the R-charged scalar, i.e., $\sRb\neq 0$. 
From the quantization condition~\eqref{Dirac_qtm:boson2}, the constant~$C_{\tht\phi}$ is determined as
\begin{align}
 \sRb C_{\tht\phi} &= \brkt{\int_0^\pi d\tht\;\frac{b(\tht)c(\tht)}{n(\tht)a^3(\tht)}e^{-\sigma(\tht)}}^{-1}\brkt{k_{\rm b}+2\sRb\tl{\xi}} \nonumber\\
 &= \frac{a_{\rm c}^3}{2r_{\rm c}b_{\rm c}^2e^{-\sigma_{\rm c}}}\brkt{k_{\rm b}+2\sRb\tl{\xi}}, 
\end{align}
where $\tl{\xi}=\gc^2\xi/\pi$. 
Thus, the background gauge potential in \eqref{expr:A_phi^pm} is calculated as
\begin{align}
 \sRb A_\phi^{(+)}(\tht) &= \frac{a_{\rm c}^3}{2r_{\rm c}b_{\rm c}^2e^{-\sigma_{\rm c}}}\brkt{k_{\rm b}+2\sRb\tl{\xi}}
 \int_0^\tht d\tht'\;\frac{r_{\rm c}b_{\rm c}^2e^{-\sigma_{\rm c}}}{a_{\rm c}^3}\sin\tht'-\sRb\tl{\xi} \nonumber\\
 &= \brkt{\frac{k_{\rm b}}{2}+\sRb\tl{\xi}}\brkt{1-\cos\tht}-\sRb\tl{\xi}. 
\end{align}
Therefore, the mode equation~\eqref{md_eq:g:q0} is
\begin{align}
 \sbk{\der_\tht^2+\cot\tht\der_\tht-\frac{\brkt{\zeta_{\rm b}-2\eta_{\rm b}\sin^2\frac{\tht}{2}}^2}{\sin^2\tht}
 +b_{\rm c}^2\lmd_q^2}f(\tht;q) &= 0, 
 \label{md_eq:rugby:boson}
\end{align}
where
\begin{align}
 \zeta_{\rm b} &\equiv \frac{1}{r_{\rm c}}\brkt{q+\sRb\tl{\xi}}, \;\;\;\;\;
 \eta_{\rm b} \equiv \frac{1}{r_{\rm c}}\brkt{\frac{k_{\rm b}}{2}+\sRb\tl{\xi}}. 
\end{align}

\begin{figure}[t]
\begin{center}
 \includegraphics[scale=0.55]{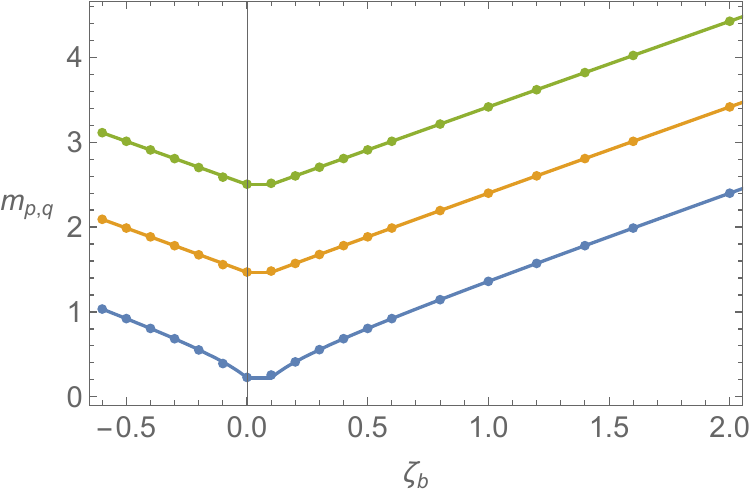}\;\;\;
 \includegraphics[scale=0.55]{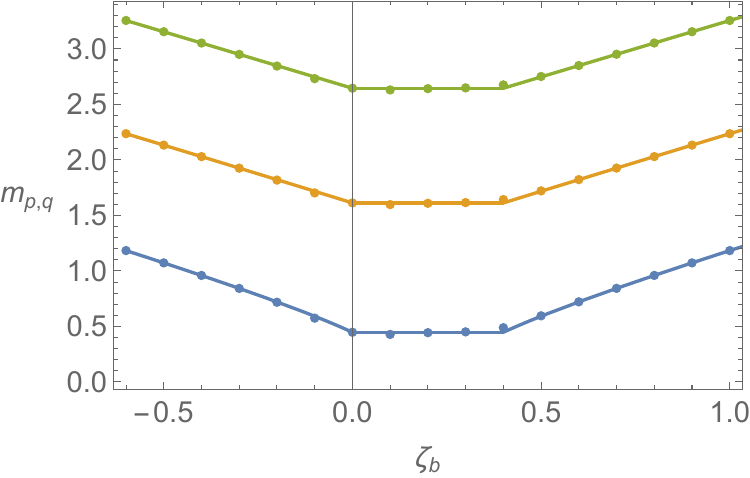}
\end{center}
 \caption{The KK mass eigenvalues~$m_{p,q}$ in the cases of $b_{\rm c}=1$, $\eta_{\rm b}=0.05$ (left plot) and $\eta_{\rm b}=0.2$ (right plot). 
 The blobs denote the values calculated by numerically solving $\cF_q(\lmd_q)=0$, where $\cF_q(\lmd_q)$ is defined in \eqref{def:cF}. 
 The lines represent \eqref{boson_KKmass} for $p=0,1,2$ from bottom to top. 
 }
\label{fig:mpq-zeta}
\end{figure}

From the numerical calculations, we find that~\footnote{
Around $\zeta_{\rm b}=0,2\eta_{\rm b}$, the derivative of the mode function~$f'(\tht;q)$ becomes very large, 
and thus the numerical error increases.   
So some careful treatment is necessary to obtain the correct values there. 
} 
\begin{align}
 m_{p,q} &= \begin{cases} \bar{\lmd}_{\rm b}(\zeta_{\rm b};p) & \brkt{\mbox{$\zeta_{\rm b} < 0$ or $\zeta_{\rm b} > 2\eta_{\rm b}$}} \\
 \sqrt{p(2\eta_{\rm b}+p)+\eta_{\rm b}+p} & \brkt{0\leq \zeta_{\rm b}\leq 2\eta_{\rm b}} \end{cases}, 
 \label{boson_KKmass}
\end{align}
where $p=0,1,2\cdots$, and 
\begin{align}
 \bar{\lmd}_{\rm b}(\zeta_{\rm b};p) &\equiv \frac{1}{b_{\rm c}}
 \sqrt{\brkt{\abs{\zeta_{\rm b}}+p}\brkt{\abs{\zeta_{\rm b}-2\eta_{\rm b}}+p+1}-{\rm sign}(\zeta_{\rm b})\eta_{\rm b}}. 
\end{align}
We have assumed that $\eta_{\rm b}\geq 0$.\footnote{
When $\eta_{\rm b}<0$, we can obtain the KK mass by applying \eqref{boson_KKmass} 
after flipping the signs of both $\zeta_{\rm b}$ and $\eta_{\rm b}$ simultaneously. 
}
Fig.~\ref{fig:mpq-zeta} shows the KK mass eigenvalues for various values of $\zeta_{\rm b}$ 
in the cases of $\eta_{\rm b}=0.05$ (left plot) and $\eta_{\rm b}=0.2$ (right plot). 
We can see that the numerical solutions of $\cF_q(\lmd_q)=0$, where $\cF_q(\lmd_q)$ is defined in \eqref{def:cF}, 
well agree with \eqref{boson_KKmass}. 
In the limit of $\sRb\to 0$ (i.e., $\eta_{\rm b}\to 0$), \eqref{boson_KKmass} reduces to \eqref{expr:barm}. 
The expression~\eqref{boson_KKmass} can be rewritten as
\begin{align}
 m_{p,q} &= \frac{1}{b_{\rm c}}\sqrt{\brkt{\frac{\abs{\zeta_{\rm b}}+\abs{\zeta_{\rm b}-2\eta_{\rm b}}}{2}+p}
 \brkt{\frac{\abs{\zeta_{\rm b}}+\abs{\zeta_{\rm b}-2\eta_{\rm b}}}{2}+p+1}-\eta_{\rm b}^2}. 
 \label{boson:m_pq}
\end{align}
This expression is applicable even when $\eta_{\rm b}<0$, 
and agrees with the result in Ref.~\cite{Williams:2012au}.\footnote{
After the change of notations, $b_{\rm c}^2m_{p,q}^2\to\lmd$, $r_{\rm c}\to\alp$, $k_{\rm b}\to N$, $p\to j$, $q\to n$, $\sRb\to q$ 
and $\sRb\tl{\xi}\to -\Phi_\pm = -\Phi/2$, we obtain (C.13) in Ref.~\cite{Williams:2012au}. 
}

\section{KK mass spectrum of bulk spinor}
\label{spinor_case}
In this section, we consider the contribution of a 6D Weyl spinor~$\Psi_+$ with + chirality as an example (i.e., $\Gm_7\Psi_+=\Psi_+$).  
The action of $\Psi_+$ is given by
\begin{align}
 S_{\rm spinor} &= \int d^6x\;\sqrt{-g^{(6)}}\: i\bar{\Psi}_+\Gm^AE_A^{\;\;M}\cD_M\Psi_+, 
 \label{S_spinor}
\end{align}
where $E_A^{\;\;M}$ is the inverse matrix of the sechsbein~$E_M^{\;\;A}$, 
and $\Gm^A$ are the 6D gamma matrices. 
The covariant derivative~$\cD_M$ is given by
\begin{align}
 \cD_M &\equiv \der_M-\frac{1}{8}\omega_M^{\;\;AB}\sbk{\Gm_A,\Gm_B}-i\sRf A_M, 
\end{align}
where $\sRf$ is the R-charge, and the spin connection~$\omega_M^{\;\;AB}$ is expressed as
\begin{align}
 \omega_M^{\;\;AB} &= E_N^{\;\;A}\brkt{\der_M E^{NB}+\Gamma^N_{\;\;LM}E^{LB}}, 
\end{align}
where $\Gm^N_{\;\;LM}$ is the affine connection.  
We should note that two coordinate patches on the compact space are necessary 
to express the sechsbein~$E_M^{\;\;A}$, just like we did for the gauge potential~$A_\phi$. 
Here we take the gauge in which the sechsbein is expressed as
\begin{align}
 E_M^{\;\;A} &= \begin{pmatrix} n & & & \\ & a\id_3 & & \\ & & b\cos\phi & \Sgm b\sin\phi \\ & & -\Sgm c\sin\phi & c\cos\phi \end{pmatrix}, 
 \label{E_M^A:gauge}
\end{align}
where $\Sgm=1$ for the patch that contain the pole~$\tht=0$, and $\Sgm=-1$ for the other patch.  
Then, after neglecting the time-dependence of the background, 
the non-vanishing components of the spin connection under our metric ansatz~\eqref{bkg:metric} are 
\begin{align}
 \omg_t^{\;\;04} &= \frac{n'}{b}\cos\phi, \;\;\;\;\;
 \omg_t^{\;\;05} = \frac{\Sgm n'}{b}\sin\phi, \nonumber\\
 \omg_i^{\;\;\udl{j}4} &= \dlt_i^{\;\;j}\frac{a'}{b}\cos\phi, \;\;\;\;\;
 \omg_i^{\;\;\udl{j}5} = \dlt_i^{\;\;j}\frac{\Sgm a'}{b}\sin\phi, \;\;\;\;\;
 \omg_\phi^{\;\;45} = \Sgm-\frac{c'}{b}. 
\end{align}
In this gauge, we find that
\begin{align}
 \lim_{\tht\to 0} \omg_\phi^{\;\;45} &= 1-r_0 = \frac{\dlt_0}{2\pi}, \nonumber\\
 \lim_{\tht\to \pi} \omg_\phi^{\;\;45} &= -1+r_\pi = -\frac{\dlt_\pi}{2\pi}, 
\end{align}
where $r_0$ and $r_\pi$ are defined in \eqref{bkg_near0} and \eqref{def:abr_pi}, 
and $\dlt_0$ and $\dlt_\pi$ are the deficit angles at $\tht=0$ and $\tht=\pi$, respectively. 
Namely, our gauge~\eqref{E_M^A:gauge} satisfies the conditions~\eqref{lim0:omg} and \eqref{limpi:omg}.\footnote{
The gauge~$E_M^A=\diag(n,a,a,a,b,c)$ does not satisfy these conditions. 
}

\subsection{Mode equations}
Since we neglect the time-dependence of the background, the Dirac equation, 
\begin{align}
 \Gm^AE_A^{\;\;M}\cD_M\Psi_+ &= 0, 
\end{align}
is expressed as
\begin{align}
 &\brkt{\frac{1}{n}\gm^0\der_t+\frac{1}{a}\gm^{\udl{i}}\der_i}\psi \nonumber\\
 &+\brkt{i\gm_5\cos\phi-\Sgm\sin\phi}
 \sbk{\frac{1}{b}\der_\tht+\frac{i}{c}\gm_5\brkt{\der_\phi-i\sRf A_\phi}
 +\frac{1}{2b}\brkt{\frac{n'}{n}+\frac{3a'}{a}+\frac{c'-\Sgm b}{c}}}\psi = 0, 
\end{align}
where $\psi$ is the 4-component spinor defined as
\begin{align}
 \Psi_+ &= \begin{pmatrix} \psi \\ \bdm{0}_4 \end{pmatrix}. 
\end{align}
In the 2-component spinor notation, this becomes
\begin{align}
 \brkt{-\frac{1}{n}\der_t+\frac{1}{a}\tau^i\der_i}\bar{\lmd}
 +e^{i\Sgm\phi}\sbk{\frac{i}{b}\der_\tht-\frac{1}{c}\brkt{\der_\phi-i\sRf A_\phi}+\frac{i}{2b}\brkt{\frac{n'}{n}+\frac{3a'}{a}+\frac{c'-\Sgm b}{c}}}\chi &= 0, \nonumber\\
 \brkt{-\frac{1}{n}\der_t-\frac{1}{a}\tau^i\der_i}\chi
 -e^{-i\Sgm\phi}\sbk{\frac{i}{b}\der_\tht+\frac{1}{c}\brkt{\der_\phi-i\sRf A_\phi}+\frac{i}{2b}\brkt{\frac{n'}{n}+\frac{3a'}{a}+\frac{c'-\Sgm b}{c}}}\bar{\lmd} &= 0, 
 \label{EOM:fermion}
\end{align}
where $\tau^i$ ($i=1,2,3$) are the Pauli matrices, and 
\begin{align}
 \psi &\equiv \begin{pmatrix} \chi \\ \bar{\lmd} \end{pmatrix}. 
\end{align}
Now we decompose the fields into the KK modes as
\begin{align}
 \chi(x^\mu,\tht,\phi) &= \sum_{p,q} \frac{e^{iq\phi}}{\sqrt{2\pi}} h_{{\rm R}p}(\tht;q)\chi_{p,q}(x^\mu), \nonumber\\
 \bar{\lmd}(x^\mu,\tht,\phi) &= \sum_{p,q} \frac{e^{iq\phi}}{\sqrt{2\pi}} h_{{\rm L}p}(\tht;q)\bar{\lmd}_{p,q}(x^\mu), 
\end{align}
where the mode functions~$h_{{\rm R}p}(\tht;q)$ and $h_{{\rm L}p}(\tht;q)$ satisfy
\begin{align}
 &\quad
 e^{i\Sgm\phi}\sbk{\frac{i}{b}\der_\tht-\frac{1}{c}\brkt{\der_\phi-i\sRf A_\phi}
 +\frac{i}{2b}\brkt{\frac{n'}{n}+\frac{3a'}{a}+\frac{c'-\Sgm b}{c}}}\brc{e^{iq\phi}h_{{\rm R}p}(\tht;q)} \nonumber\\ 
 &= \frac{im_{p,q}}{n}\brc{e^{i(q+\Sgm)\phi}h_{{\rm L}p}(\tht;q+\Sgm)}, \nonumber\\
 &\quad
 e^{-i\Sgm\phi}\sbk{\frac{i}{b}\der_\tht+\frac{1}{c}\brkt{\der_\phi-i\sRf A_\phi}
 +\frac{i}{2b}\brkt{\frac{n'}{n}+\frac{3a'}{a}+\frac{c'-\Sgm b}{c}}}\brc{e^{i(q+\Sgm)\phi}h_{{\rm L}p}(\tht;q+\Sgm)} \nonumber\\
 &= -\frac{im_{p,q}}{n}\brc{e^{iq\phi}h_{{\rm R}p}(\tht;q)}. 
 \label{md_eq:fermion}
\end{align}
Then, \eqref{EOM:fermion} is rewritten as
\begin{align}
 \sum_{p,q}\frac{e^{i(q+\Sgm)\phi}}{\sqrt{2\pi}}h_{{\rm L}p}(\tht;q+\Sgm)\sbk{\brkt{-\frac{1}{n}\der_t+\frac{1}{a}\tau^i\der_i}\bar{\lmd}_{p,q+\Sgm}(x^\mu)
 +\frac{im_{p,q}}{n}\chi_{p,q}(x^\mu)} &= 0, \nonumber\\
 \sum_{p,q}\frac{e^{iq\phi}}{\sqrt{2\pi}}h_{{\rm R}p}(\tht;q)\sbk{\brkt{-\frac{1}{n}\der_t-\frac{1}{a}\tau^i\der_i}\chi_{p,q}(x^\mu)
 +\frac{im_{p,q}}{n}\bar{\lmd}_{p,q+\Sgm}(x^\mu)} &= 0. 
\end{align}
Using the rescaling of $x^i$ in \eqref{rescaling} and the orthonormal relation, 
\begin{align}
 \int_0^\pi d\tht\,a^3(\tht)b(\tht)c(\tht)h_{{\rm R}p}(\tht;q)h_{{\rm R}p'}(\tht;q) 
 &= \dlt_{p,p'}, \nonumber\\
 \int_0^\pi d\tht\,a^3(\tht)b(\tht)c(\tht)h_{{\rm L}p}(\tht;q)h_{{\rm L}p'}(\tht;q) 
 &= \dlt_{p,p'}, 
\end{align}
we can pick up the equations of motion for each KK mode. 
\begin{align}
 \brkt{-\der_t+\tau^i\der_{\tl{x}^i}}\bar{\lmd}_{p,q+\Sgm}(x^\mu)+im_{p,q}\chi_{p,q}(x^\mu) &= 0, \nonumber\\
 \brkt{-\der_t-\tau^i\der_{\tl{x}^i}}\chi_{p,q}(x^\mu)+im_{p,q}\bar{\lmd}_{p,q+\Sgm}(x^\mu) &= 0. 
\end{align}
Hence we have the same dispersion relation as \eqref{dispersion_rel} for the fermions. 
Notice that $\chi_{p,q}(x^\mu)$ forms the mass term with $\bar{\lmd}_{p,q+\Sgm}(x^\mu)$, 
rather than $\bar{\lmd}_{p,q}(x^\mu)$.

\subsection{Mass eigenvalues}
The mode equations in \eqref{md_eq:fermion} are rewritten as 
\begin{align}
 \sbk{\der_\tht-\frac{b}{c}\brkt{q_\Sgm-\sRf A_\phi}}\tl{u}_{{\rm R}p}(\tht;q) &= \frac{bm_{p,q}}{n}\tl{h}_{{\rm L}p}(\tht;q+\Sgm), \nonumber\\
 \sbk{\der_\tht+\frac{b}{c}\brkt{q_\Sgm-\sRf A_\phi}}\tl{u}_{{\rm L}p}(\tht;q+\Sgm) &= -\frac{bm_{p,q}}{n}\tl{h}_{{\rm R}p}(\tht;q), 
 \label{md_eq:fm2}
\end{align}
where
\begin{align}
 q_\Sgm &\equiv q+\frac{\Sgm}{2}, \nonumber\\
 \tl{h}_{{\rm R}p}(\tht;q) &\equiv \sqrt{n(\tht)a^3(\tht)c(\tht)}h_{{\rm R}p}(\tht;q), \nonumber\\
 \tl{h}_{{\rm L}p}(\tht;q) &\equiv \sqrt{n(\tht)a^3(\tht)c(\tht)}h_{{\rm L}p}(\tht;q). 
\end{align}
We can further rewrite the above equations as
\begin{align}
 &\quad
 \left[\der_\tht^2+\brkt{\frac{n'}{n}-\frac{b'}{b}}\der_\tht-\frac{b}{c}\brkt{\frac{n'}{n}-\frac{c'}{c}}\brkt{q_\Sgm-\sRf A_\phi} \right.\nonumber\\
 &\hspace{7mm}\left.
 +\frac{b}{c}\sRf A_\phi'+\frac{b^2m_{p,q}^2}{n^2}-\frac{b^2}{c^2}\brkt{q_\Sgm-\sRf A_\phi}^2\right]\tl{h}_{{\rm R}p}(\tht;q) = 0, \nonumber\\
 &\quad
 \left[\der_\tht^2+\brkt{\frac{n'}{n}-\frac{b'}{b}}\der_\tht+\frac{b}{c}\brkt{\frac{n'}{n}-\frac{c'}{c}}\brkt{q_\Sgm-\sRf A_\phi} \right.\nonumber\\
 &\hspace{7mm}\left.
 -\frac{b}{c}\sRf A'_\phi+\frac{b^2m_{p,q}^2}{n^2}-\frac{b^2}{c^2}\brkt{q_\Sgm-\sRf A_\phi}^2\right]\tl{h}_{{\rm L}p}(\tht;q+\Sgm) = 0. 
\end{align}
These equations do not change their forms in both patches since $q_\Sgm-\sRf A_\phi^{(\Sgm)}$ is gauge-invariant. 
Thus, we express them as 
\begin{align}
 \sbk{\der_\tht^2+\brkt{\frac{n'}{n}-\frac{b'}{b}}\der_\tht-\frac{b}{c}\brkt{\frac{n'}{n}-\frac{c'}{c}}\cD_q
 -\frac{b}{c}\cD_q'+\frac{b^2\lmd_q^2}{n^2}-\frac{b^2}{c^2}\cD_q^2}\tl{h}_{\rm R}(\tht;q) &= 0, \nonumber\\
 \sbk{\der_\tht^2+\brkt{\frac{n'}{n}-\frac{b'}{b}}\der_\tht+\frac{b}{c}\brkt{\frac{n'}{n}-\frac{c'}{c}}\cD_q
 +\frac{b}{c}\cD_q'+\frac{b^2\lmd_q^2}{n^2}-\frac{b^2}{c^2}\cD_q^2}\tl{h}_{\rm L}(\tht;q+1) &= 0, 
 \label{md_eq:tlh}
\end{align}
where
\begin{align}
 \cD_q(\tht) &\equiv q_{+1}-\sRf A_\phi^{(+)}(\tht) = q+\frac{1}{2}-\sRf A_\phi^{(+)}(\tht). 
 \label{def:cD_q}
\end{align}
Since $\tl{h}_{\rm R}(\tht;q)$ and $\tl{h}_{\rm L}(\tht;q+1)$ share the same eigenvalue~$\lmd_q$, 
it is enough to focus on the first equation in \eqref{md_eq:tlh} in order to evaluate the KK mass spectrum.

The behaviors near the poles~$\tht=0,\pi$ are collected in Appendix~\ref{behaviors:spinor}. 
From the behavior near $\tht=0$, we can choose the initial conditions as
\begin{align}
 \tl{h}_{\rm R}(\ep;q) &= \ep^{\zeta_{\rm f}}, \;\;\;\;\;
 \tl{h}'_{\rm R}(\ep;q) = \zeta_{\rm f}\ep^{\zeta_{\rm f}-1}, \;\;\;\;\;
 \brkt{\zeta_{\rm f} \equiv \frac{1}{r_0}\brkt{q+\frac{1}{2}+\sRf \tl{\xi}}}
 \label{ini_cond:1}
\end{align}
for $0\leq \zeta_{\rm f}\leq \frac{1}{2}$ or $\zeta_{\rm f}>1$, and 
\begin{align}
 \tl{h}_{\rm R}(\ep;q) &= \ep^{1-\zeta_{\rm f}}, \;\;\;\;\;
 \tl{h}'_{\rm R}(\ep;q) = \brkt{1-\zeta_{\rm f}}\ep^{-\zeta_{\rm f}}, 
 \label{ini_cond:2}
\end{align}
for $\zeta_{\rm f}<0$ or $\frac{1}{2}<\zeta_{\rm f}\leq 1$. 
The constant~$r_0$ in the definition of $\zeta_{\rm f}$ is defined in \eqref{bkg_near0}. 
From the behavior near $\tht=\pi$, we define
\begin{align}
 \cF_q(\lmd_q) &= \begin{cases} \displaystyle \lim_{\tht\to\pi}\brkt{\pi-\tht}^{-1-\alp_{\rm f}}\tl{h}_{\rm R}(\tht;q) 
 & \displaystyle \brkt{\alp_{\rm f} < -\frac{1}{2}} \\
 \displaystyle \lim_{\tht\to\pi}\brkt{\pi-\tht}^{\alp_{\rm f}}\tl{h}_{\rm R}(\tht;q) & \displaystyle \brkt{\alp_{\rm f}\geq -\frac{1}{2}} \end{cases}, 
 \label{def:cF:2}
\end{align}
where
\begin{align}
 \alp_{\rm f} &\equiv \frac{1}{r_\pi}\brkt{q+\frac{1}{2}-k_{\rm f}-\sRf \tl{\xi}}. 
\end{align}
The constant~$r_\pi$ is defined in \eqref{def:abr_pi}. 
Then, the KK masses~$\lmd_q=m_{p,q}$ ($p=0,1,2,\cdots$) are obtained as solutions of $\cF_q(\lmd_q)=0$.

\subsection{Rugby-ball background}
For the rugby-ball background~\eqref{bkg:simple_case}, $\cD_q(\tht)$ defined in \eqref{def:cD_q} becomes 
\begin{align}
 \cD_q(\tht) &= q+\frac{1}{2}+\sRf\tl{\xi}-\brkt{\frac{k_{\rm f}}{2}+\sRf\tl{\xi}}\brkt{1-\cos\tht}. 
\end{align}
We have used \eqref{expr:A_phi^pm} and \eqref{Dirac_qtm:fermion}. 
Thus, the mode equation~\eqref{md_eq:tlh} is reduced to 
\begin{align}
 \sbk{\der_\tht^2+\frac{\gm_{\rm f}\cos\tht+\eta_{\rm f}-\brkt{\gm_{\rm f}+\eta_{\rm f}\cos\tht}^2}{\sin^2\tht}+b_{\rm c}^2\lmd_q^2}
 \tl{h}_{\rm R}(\tht;q) &= 0, 
 \label{md_eq:fm:rugby}
\end{align}
where
\begin{align}
 \gm_{\rm f} &\equiv \frac{1}{r_{\rm c}}\brkt{q+\frac{1}{2}-\frac{k_{\rm f}}{2}}, \;\;\;\;\;
 \eta_{\rm f} \equiv \frac{1}{r_{\rm c}}\brkt{\frac{k_{\rm f}}{2}+\sRf\tl{\xi}}. 
\end{align}
By numerically integrating this equation with the initial conditions in \eqref{ini_cond:1} or \eqref{ini_cond:2}, 
we find that the KK mass eigenvalues are~\footnote{
We have assumed that $\eta_{\rm f}\geq 0$. 
} 
\begin{align}
 m_{p,q} &= \begin{cases} \displaystyle 
 \frac{1}{b_{\rm c}}\sqrt{\brkt{\abs{\gm_{\rm f}+\eta_{\rm f}}+p+\frac{1}{2}}\brkt{\abs{\gm_{\rm f}-\eta_{\rm f}}+p+\frac{1}{2}}} 
 & \displaystyle \brkt{\abs{\gm_{\rm f}} \geq \eta_{\rm f}+\frac{1}{2}} \\
 \displaystyle \frac{1}{b_{\rm c}}\sqrt{(p+1)\brkt{2\abs{\eta_{\rm f}}+p+1}} & \displaystyle \brkt{\abs{\gm_{\rm f}} < \eta_{\rm f}+\frac{1}{2}} \end{cases}, 
\end{align}
where $p=0,1,2,\cdots$.  
This can be rewritten as
\begin{align}
 m_{p,q} &= \frac{1}{b_{\rm c}}\sqrt{\brkt{p+\frac{1}{2}\abs{\gm_{\rm f}+\eta_{\rm f}+\frac{1}{2}}
 +\frac{1}{2}\abs{\gm_{\rm f}-\eta_{\rm f}-\frac{1}{2}}+\frac{1}{2}}^2-\eta_{\rm f}^2}, 
 \label{fermion:m_pq}
\end{align}
which agrees with the result in Ref.~\cite{Williams:2012au}. 
The expression~\eqref{fermion:m_pq} is applicable even when $\eta_{\rm f}<0$.

We should note that the deficit angle~$\dlt_{\rm df}$ is related to the brane tension~$\tau$ as $\dlt_{\rm df}=\tau$~\cite{Vinet:2004bk}, 
and the SUSY condition for the background relates $\tau$ and the FI parameter~$\xi$ as $\tau=C_{\tau\xi}\tl{\xi}$, where $C_{\tau\xi}$ is a constant. 
Thus, from \eqref{rel:dlt-rc}, $\tl{\xi}$ is expressed as
\begin{align}
 \tl{\xi} &= \frac{\tau}{C_{\tau\xi}} = \frac{\dlt_{\rm df}}{C_{\tau\xi}} = \frac{2\pi}{C_{\tau\xi}}\brkt{1-r_{\rm c}}. 
\end{align}
Comparing the scalar mass~\eqref{boson:m_pq} and the spinor mass~\eqref{fermion:m_pq} with this relation, they are degenerate iff
\begin{align}
 C_{\tau\xi} &= -\frac{4\pi}{k}, \;\;\;\;\;
 s_{\rm b} = 0, \;\;\;\;\;
 s_{\rm f} = -\frac{1}{k}, 
 \label{SUSY_cond}
\end{align}
where $k$ is defined as $k_{\rm b}\equiv s_{\rm b}k$ (and also $k_{\rm f}=s_{\rm f}k$). 
This corresponds to the SUSY condition. 
In fact, plugging these relations into \eqref{boson:m_pq} and \eqref{fermion:m_pq}, they become
\begin{align}
 m_{p,q} &= \frac{1}{b_{\rm c}}\sqrt{\brkt{p+\frac{\abs{q}}{r_{\rm c}}}\brkt{p+\frac{\abs{q}}{r_{\rm c}}+1}}, \;\;\;\;\;
 \brkt{\mbox{scalar}} \nonumber\\
 m_{p,q'} &= \frac{1}{b_{\rm c}}\sqrt{\brkt{p+\frac{\abs{q'+1}}{r_{\rm c}}}\brkt{p+\frac{\abs{q'+1}}{r_{\rm c}}+1}}, \;\;\;\;\;
 \brkt{\mbox{spinor}}
 \label{expr:m_pq:SUSY}
\end{align}
where $p=0,1,2,\cdots$ and $q,q'\in{\mathbb Z}$. 
Therefore, the two spectra agree in this case.

\subsection{Mass-splitting due to nontrivial $\theta$-dependence of background}
Here we perturb the (SUSY) rugby-ball solution~\eqref{bkg:simple_case} by adding small $\tht$-dependent terms, 
and see how the KK mass spectrum is modified. 

\begin{figure}[t]
\begin{center}
 \includegraphics[scale=0.55]{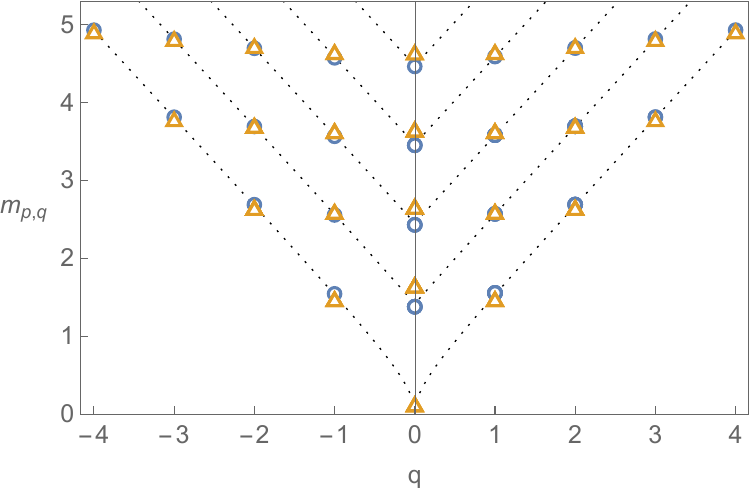}\;\;\;
 \includegraphics[scale=0.55]{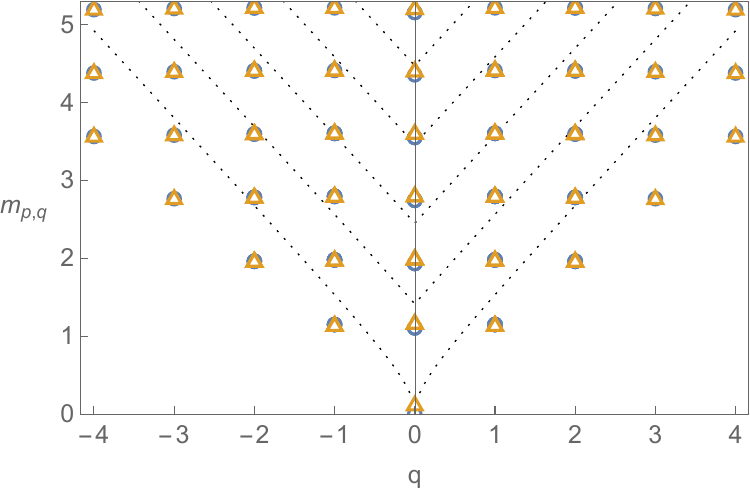}
\end{center}
 \caption{The KK mass eigenvalues~$m_{p,q}$ in the case of \eqref{a-deviate} (left plot) and that of \eqref{b-deviate} (right plot). 
 The (blue) circles and the (orange) triangles correspond to the spinor and the scalar KK masses, respectively. 
 The dotted lines denote the scalar KK masses in the supersymmetric case~\eqref{expr:m_pq:SUSY} 
 with $b_{\rm c}=1$, $\sgm_{\rm c}=1$ and $r_{\rm c}=0.9$. 
 }
\label{fig:SUSY-breaking}
\end{figure}

First we consider the case that the 3D scale factor~$a(\tht)$ deviates from \eqref{bkg:simple_case} as 
\begin{align}
 a(\tht) &= a_{\rm c} \;\;\; \to \;\;\; a_{\rm c}\sbk{1+0.4\sin\tht}. 
 \label{a-deviate}
\end{align}
The corresponding KK mass eigenvalues are shown by the left plot in Fig.~\ref{fig:SUSY-breaking}.~\footnote{
We plot $m_{p,q-1}$ in order to match the bosonic spectrum. 
}  
We can see that both the deviation from \eqref{expr:m_pq:SUSY} and the mass splittings between the bosons and the fermions 
are small, but the lighter $q=0$ modes receive relatively larger SUSY-breaking effect. 

Next we perturb the extra-dimensional scale factors~$b(\tht)$ and $c(\tht)$ as 
\begin{align}
 b(\tht) &= b_{\rm c} \;\;\; \to \;\;\; b_{\rm c}\sbk{1+0.4\sin\tht}, \nonumber\\
 c(\tht) &= r_{\rm c}b_{\rm c}\sin\tht \;\;\; \to \;\;\; r_{\rm c}b(\tht)\sin\tht. 
 \label{b-deviate}
\end{align}
In this case, as shown in the right plot of Fig.~\ref{fig:SUSY-breaking}, 
but the SUSY-breaking mass-splitting is much smaller than the deviation from \eqref{expr:m_pq:SUSY}.

\section{Summary}
\label{summary}
We derived the mode equations for a bulk scalar and a bulk spinor, 
including the lapse function and the 3D scale factor, and evaluate the KK masses for backgrounds 
that have nontrivial dependence on the position in the compact space. 
In particular, we consider several perturbations of the rugby-ball background in 6D SUGRA~\cite{Carroll:2003db,Aghababaie:2003wz,Burgess:2004ib}. 
By integrating the mode equations, we obtain the function~$\cF_q(\lmd_q)$ whose zeros are the KK mass eigenvalues. 
Roughly, the KK masses are determined by the volume of the compact space. 
However, when the 3D scale factor (or the lapse function) depends on the extra-dimensional coordinates~$y^m$ ($m=1,2$), 
it also affects the KK masses. 
In fact, after the moduli stabilization, the scale factors generically depend on $y^m$~\cite{Goldberger:1999uk,Burgess:2012pc}. 
In order to understand the effects of various background deformations on the spectrum, 
we numerically evaluate the KK masses by solving $\cF_q(\lmd_q)=0$, and obtain the following results. 
\begin{itemize}
 \item The lower KK modes are less sensitive to a nontrivial $\tht$-dependence of the extra-dimensional scale factor~$b$ (or $c$). 
 
 \item The effect of the 3D scale factor~$a$ on the spectrum is small compared to that of the extra-dimensional one, 
 but the lower KK modes may receive a non-negligible contribution. 
 
 \item From the mass formulae, the SUSY background only allows R-neutral hypermultiplets (i.e., $\sRb=0$).  
 
 \item For a deformation of the (SUSY) rugby-ball background, 
 the mass-splitting between a boson and a fermion is much smaller than the deviation from \eqref{b-deviate}. 

\end{itemize}
In the evaluation of the KK masses~$m_{p,q}$, we have dropped the $t$-dependence of the background. 
In the cosmological evolution, the lapse function~$n$ and the scale factors~$a,b,c$ change from moment to moment. 
Thus, the time dependence of $m_{p,q}$ is taken into account through the time evolution of the functions~$\{n,a,b,c\}$ 
in the mode equations. 

The function~$\cF_q(\lmd_q)$ defined in \eqref{def:cF} or \eqref{def:cF:2} is also useful 
when we sum up all the contributions of the KK modes by using the analytic continuation and the contour integral for the complex plane~\cite{Garriga:2000jb}. 
In fact, the energy density and the pressure of the radiation in the early universe, or the Casimir energy for the compact space can be 
calculated by such a method. 
Those quantities play important roles in cosmological discussions. 
In our subsequent works, we will investigate the scale-factor dependencies on them, 
and pursue their time evolutions by numerically solving the 6D Einstein equations. 

\section*{Acknowledgments}

This work was supported by JSPS KAKENHI Grant Number JP25H01539 (H.O.).

\appendix

\section{Flux quantization} \label{flux_qtz}
From \eqref{6Daction} and \eqref{cL_brane}, the Lagrangian terms involving the field strength are~\cite{Lee:2007vy}
\begin{align}
 {\cal L}_{\rm FF} &= \sqrt{-g^{(6)}}\sbk{-\frac{e^\sigma}{4\gc^2}F^{MN}F_{MN}
 +\xi_+ e^\sigma\epsilon^{mn}F_{mn}\delta^{(2)}(y-y_+)
 +\xi_- e^\sigma\epsilon^{mn}F_{mn}\delta^{(2)}(y-y_-)} \nonumber\\
 &= \sqrt{-g^{(6)}}\sbk{-\frac{e^\sigma}{2\gc^2b^2c^2}F_{\tht\phi}^2
 +\frac{2\xi_+ e^\sigma}{bc}F_{\tht\phi}\dlt^{(2)}(y-y_+)
 +\frac{2\xi_- e^\sigma}{bc}F_{\tht\phi}\dlt^{(2)}(y-y_-)+\cdots}. 
 \label{cL_FF}
\end{align}
From \eqref{expr:F_thtphi}, the field strength~$F_{\tht\phi}$ has the following profile in the bulk. 
\begin{align}
 F^{\rm (bulk)}_{\tht\phi}(\tht) &= C_{\tht\phi}\frac{b(\tht)c(\tht)}{n(\tht)a^3(\tht)}e^{-\sigma(\tht)}. 
\end{align}
Due to the brane-localized flux terms, $F_{\tht\phi}$ also has the contributions proportional to the delta functions. 
From \eqref{cL_FF},  they are identified as 
\begin{align}
 F_{\tht\phi} &\equiv F^{\rm (bulk)}_{\tht\phi}-2\gc^2bc\sbk{\xi_+\dlt^{(2)}(y-y_+)+\xi_-\dlt^{(2)}(y-y_-)}. 
 \label{expr:hatF}
\end{align}
Here, the 2D delta function~$\dlt^{(2)}(y-y_\pm)$ is defined so that
\begin{align}
 \int_{\cA_\pm}d^2y\;\sqrt{\det(g_{mn})}\dlt^{(2)}(y-y_\pm) &= 1, 
 \label{def:delta}
\end{align}
where $\cA_\pm$ is a region that contains $y=y_\pm$, and the anti-symmetric tensor~$\epsilon^{mn}$ is defined as 
\begin{align}
 \epsilon^{12} &= -\epsilon^{21} = \frac{1}{\sqrt{\det(g_{mn})}} = \frac{1}{bc}, \nonumber\\
 \epsilon^{11} &= \epsilon^{22} = 0. 
 \label{def:epsilon}
\end{align}

Here let us consider a region~$\cA_+=\brc{(\tht,\phi)|0\leq \tht\leq \tht_*, 0\leq \phi<2\pi}$, 
where $\tht_*$ is a constant smaller than $\pi$. 
From the Stokes theorem, we have
\begin{align}
 \int_0^{2\pi}d\phi\;A_\phi^{(+)}(\tht_*) &= \int_{\cA_+} d^2y\;F_{\tht\phi}
\end{align}
Since $A_\phi^{(+)}(\tht_*)$ is independent of $\phi$, the left-hand-side is $2\pi A_\phi^{(+)}(\tht_*)$. 
Using \eqref{expr:hatF}, the right-hand-side is
\begin{align}
 \mbox{(RHS)} &= \int_{\cA_+}d^2y\;F_{\tht\phi}^{\rm (bulk)}-2\gc^2\xi_+
 = 2\pi\int_0^{\tht_*}d\tht\;C_{\tht\phi}\frac{b(\tht)c(\tht)}{n(\tht)a^3(\tht)}e^{-\sigma(\tht)}-2\gc^2\xi_+. 
\end{align}
We have used \eqref{expr:F_thtphi} in the second equation. 
Therefore, we obtain
\begin{align}
 A_\phi^{(+)}(\tht_*) &= C_{\tht\phi}\int_0^{\tht_*}d\tht\;\frac{b(\tht)c(\tht)}{n(\tht)a^3(\tht)}e^{-\sigma(\tht)}-\frac{\gc^2\xi_+}{\pi}. 
 \label{expr:A^+}
\end{align}
Similarly, for a region~$\cA_-\equiv S^2\setminus \cA_+=\brc{(\tht,\phi)|\tht_*< \tht\leq \pi, 0\leq \phi<2\pi}$, 
we obtain
\begin{align}
 A_\phi^{(-)}(\tht_*) &= -C_{\tht\phi}\int_{\tht_*}^\pi d\tht\;\frac{b(\tht)c(\tht)}{n(\tht)a^3(\tht)}e^{-\sigma(\tht)}+\frac{\gc^2\xi_-}{\pi}. 
 \label{expr:A^-}
\end{align}
Note that $A^{(+)}_\phi$ and $A^{(-)}_\phi$ are related through the R gauge transformation, 
\begin{align}
 \Phi^{(+)}(x) &= \exp\brkt{i\sRb\vartheta(x)}\Phi^{(-)}(x), \nonumber\\
 A_\phi^{(+)}(x) &= A_\phi^{(-)}(x)+\der_\phi\vartheta(x). 
 \label{transition}
\end{align}
From \eqref{expr:A^+} and \eqref{expr:A^-}, we have
\begin{align}
 \der_\phi\vartheta &= A_\phi^{(+)}-A_\phi^{(-)} \nonumber\\
 &= C_{\tht\phi}\int_0^\pi d\tht\;\frac{b(\tht)c(\tht)}{n(\tht)a^3(\tht)}e^{-\sigma(\tht)}-\frac{\gc^2}{\pi}\brkt{\xi_++\xi_-}, 
 \label{dervartht}
\end{align}
which is a constant. 
Therefore, the transition function~$\vartheta(x)$ is identified as
\begin{align}
 \vartheta(\phi) &= \phi\sbk{C_{\tht\phi}\int_0^\pi d\tht\;\frac{b(\tht)c(\tht)}{n(\tht)a^3(\tht)}e^{-\sigma(\tht)}
 -\frac{\gc^2}{\pi}\brkt{\xi_++\xi_-}}+\mbox{(constant)}. 
\end{align}
From the single-valuedness of a scalar field~$\Phi$ in the overlapped region of the coordinate patches, we obtain the quantization condition, 
\begin{align}
 \sRb C_{\rm \tht\phi}\int_0^\pi d\tht\;\frac{b(\tht)c(\tht)}{n(\tht)a^3(\tht)}e^{-\sigma(\tht)}
 -\frac{\sRb\gc^2}{\pi}\brkt{\xi_++\xi_-}  
 &= k_{\rm b}, 
 \label{Dirac_qtm:boson}
\end{align}
where $k_{\rm b}\in{\mathbb Z}$, and from the single-valuedness of a spinor field~$\Psi_+$, we obtain
\begin{align}
  \sRf C_{\rm \tht\phi}\int_0^\pi d\tht\;\frac{b(\tht)c(\tht)}{n(\tht)a^3(\tht)}e^{-\sigma(\tht)}
 -\frac{\sRf\gc^2}{\pi}\brkt{\xi_++\xi_-}  
 &= k_{\rm f}, 
 \label{Dirac_qtm:fermion}
\end{align}
where $k_{\rm f}\in{\mathbb Z}$. 
Therefore, $k_{\rm b}$ and $k_{\rm f}$ are related as $\sRf k_{\rm b}=\sRb k_{\rm f}$, 
which indicates that the ratio of $\sRb/\sRf$ is a rational number.

\section{Pole values of spin connection}
\label{expr:sechsbein}
Here we focus on the compact space. 
Near $\tht=0$, the 2D internal metric is 
\begin{align}
 ds_2^2 &= b^2(\tht)d\tht^2+c^2(\tht)d\phi^2 
 \simeq b_0^2\brkt{d\tht^2+r_0^2\tht^2d\phi^2}. 
\end{align}
The deficit angle~$\dlt_0$ at $\tht=0$ is given by $\dlt_0=2\pi(1-r_0)$. 
This conical singularity induces the 2D delta function contribution to the Ricci scalar~$R$. 
In fact, for an arbitrary function~$f(\vec{y})$ ($\vec{y}$ are the 2D coordinates), we have~\cite{Parameswaran:2006db}
\begin{align}
 \int d^2y\;\sqrt{g^{(2)}}Rf(\vec{y}) &= 2\dlt_0 f(\vec{0})+\cdots, 
 \label{int:singular}
\end{align}
where the ellipsis denotes contributions from the regular part of $R$, and will vanish when the integration region is infinitely small region 
that contains the pole~$\tht=0$. 
Here notice that 
\begin{align}
 R &= E_a^{\;\;m}E_b^{\;\;n}R_{mn}^{\;\;\;\;\;\;ab} = \frac{2}{bc}R_{\tht\phi}^{\;\;\;\;45}, 
\end{align}
where $m,n=\tht,\phi$ are the coordinate indices, and $a,b=4,5$ are the flat indices. 
Hence, from \eqref{int:singular}, we have
\begin{align}
 \int_{\cA_\epsilon} d^2y\;R_{\tht\phi}^{\;\;\;\;45} &= \dlt_0. 
\end{align}
We have chosen the integration region as $\cA_\ep=\{(\tht,\phi)|0\leq \tht\leq \ep\}$, where $\ep$ is a positive infinitesimal number. 
Since 
\begin{align}
 R_{\tht\phi}^{\;\;\;\;45} &= \der_\tht\omg_\phi^{\;\;45}-\der_\phi\omg_\tht^{\;\;45}+\cdots, 
\end{align}
the following Stokes theorem holds.  
\begin{align}
 \int_S d\tht\wedge d\phi\;R_{\tht\phi}^{\;\;\;\;45} &= \oint_C d\phi\;\omg_\phi^{\;\;45}, 
\end{align}
where $C$ is the boundary of $\cA_\ep$. 
Therefore, we obtain
\begin{align}
 \lim_{\tht\to 0}\omg_\phi^{\;\;45} &= \frac{\dlt_0}{2\pi}. 
 \label{lim0:omg}
\end{align}

Similarly, we also have
\begin{align}
 \lim_{\tht\to\pi}\omg_\phi^{\;\;45} &= -\frac{\dlt_\pi}{2\pi}, 
 \label{limpi:omg}
\end{align} 
where $\dlt_\pi$ is the deficit angle at $\tht=\pi$.

\section{Behaviors of mode functions near the poles}
\label{behaviors_near_poles}
\subsection{Scalar case} \label{behaviors:scalar}
The background fields behave near $\tht=0$ as
\begin{align}
 n(\tht) &\simeq n_0, \;\;\;
 a(\tht) \simeq a_0, \;\;\;
 b(\tht) \simeq b_0, \;\;\;
 c(\tht) \simeq r_0b_0\tht, \;\;\;
 \sigma(\tht) \simeq \sigma_0, 
 \label{bkg_near0}
\end{align}
where $n_0$, $a_0$, $b_0$, $r_0$ and $\sigma_0$ are constants. 
Then, from \eqref{expr:A_phi^pm}, we have
\begin{align}
 q-\sRb A_\phi^{(+)}(\tht) &\simeq -\sRb r_0 D_0\tht^2+r_0\zeta_{\rm b},  
\end{align}
where 
\begin{align}
 D_0 &\equiv \frac{C_{\tht\phi}b_0^2e^{-\sgm_0}}{2n_0a_0^3}, \;\;\;\;\;
 \zeta_{\rm b} \equiv \frac{q+\sRb\tl{\xi}}{r_0}, \;\;\;\;\;
 \tl{\xi} \equiv \frac{\gc^2\xi}{\pi}. 
 \label{def:zeta_b}
\end{align} 
Thus, the mode equation~\eqref{md_eq:g:q0} becomes
\begin{align}
 \sbk{\der_\tht^2+\frac{1}{\tht}\der_\tht-\frac{\zeta_{\rm b}^2}{\tht^2}+2\sRb D_0\zeta_{\rm b}+\tl{\lmd}_q^2}f(\tht;q) &\simeq 0, 
\end{align}
where $\tl{\lmd}_q\equiv b_0\lmd_q/n_0$. 
When $q\neq -\sRb\tl{\xi}$, the constant terms in the square bracket can be neglected, and the solution is
\begin{align}
 f(\tht;q) &\simeq c_1\tht^{\abs{\zeta_{\rm b}}}+c_2\tht^{-\abs{\zeta_{\rm b}}}, 
\end{align}
where $c_1$ and $c_2$ are integration constants. 
Demanding that $f(\tht;q)$ is finite at $\tht=0$, we find that $c_2=0$ and thus 
$f(\tht;q)= \cO\brkt{\tht^{\abs{\zeta_{\rm b}}}}$. 
Notice that $f'(\tht;q)$ diverges at $\tht=0$ when $\abs{\zeta_{\rm b}}<1$. 
When $q=-\sRb\tl{\xi}$, the solution becomes 
\begin{align}
 f(\tht;q) &\simeq c_1J_0\brkt{\tl{\lmd}_q\tht}+c_2Y_0\brkt{\tl{\lmd}_q\tht}. 
\end{align}
The finiteness at $\tht=0$ requires that $c_2=0$. 
Thus, $f(\tht;q)$ is almost constant near $\tht=0$. 
In summary, the behavior of the mode function near $\tht=0$ is 
\begin{align}
 f(\tht;q) &\simeq \cO\brkt{\tht^{\abs{\zeta_{\rm b}}}}. 
\end{align}

Next we consider the behavior near $\tht=\pi$. 
The background fields behave as
\begin{align}
 n(\tht) &\simeq n_\pi, \;\;\;
 a(\tht) \simeq a_\pi, \;\;\;
 b(\tht) \simeq b_\pi, \;\;\;
 c(\tht) \simeq r_\pi b_\pi(\pi-\tht), \;\;\;
 \sigma(\tht) \simeq \sigma_\pi, 
 \label{def:abr_pi}
\end{align}
where $n_\pi$, $a_\pi$, $b_\pi$, $r_\pi$ and $\sigma_\pi$ are constants. 
Then, from \eqref{expr:A_phi^pm}, \eqref{dervartht} and \eqref{Dirac_qtm:boson}, we have
\begin{align}
 q-\sRb A_\phi^{(+)}(\tht) &= q-\sRb A_\phi^{(-)}(\tht)-k_{\rm b} \nonumber\\
 &\simeq \sRb r_\pi D_\pi(\pi-\tht)^2+r_\pi \alp_{\rm b},  
\end{align}
where 
\begin{align}
 D_\pi &\equiv \frac{C_{\tht\phi}b_\pi^2 e^{-\sgm_\pi}}{2n_\pi a_\pi^3}, \;\;\;\;\;
 \alp_{\rm b} \equiv \frac{q-k_{\rm b}-\sRb\tl{\xi}}{r_\pi}. 
 \label{def:alp_b}
\end{align}
Thus, \eqref{md_eq:g:q0} becomes
\begin{align}
 \sbk{\der_\tht^2-\frac{1}{\pi-\tht}\der_\tht-\frac{\alp_{\rm b}^2}{\brkt{\pi-\tht}^2}-2\sRb D_\pi\alp_{\rm b}+\hat{\lmd}_q^2}f(\tht;q) &\simeq 0, 
\end{align}
where $\hat{\lmd}_q\equiv b_\pi\lmd_q/n_\pi$. 
When $q\neq \sRb\tl{\xi}+k_{\rm b}$, the constant terms in the square bracket can be neglected, and the solution is
\begin{align}
 f(\tht;q) &\simeq c_1\brkt{\pi-\tht}^{\alp_{\rm b}}+c_2\brkt{\pi-\tht}^{-\alp_{\rm b}}, 
\end{align}
where $c_1$ and $c_2$ are integration constants. 
When $q=\sRb\tl{\xi}+k_{\rm b}$, the solution becomes
\begin{align}
 f(\tht;q) &\simeq c_1J_0\brkt{\hat{\lmd}_q(\pi-\tht)}+c_2Y_0\brkt{\hat{\lmd}_q(\pi-\tht)} \nonumber\\
 &\simeq \mbox{constant}. 
\end{align}

In summary, $f(\tht;q)$ behaves near $\tht=\pi$ as
\begin{align}
 f\brkt{\tht;q} &\simeq \cO\brkt{\brkt{\pi-\tht}^{-\abs{\alp_{\rm b}}}}. 
 \label{g:diverge}
\end{align}

\subsection{Spinor case} \label{behaviors:spinor}
We first consider the behaviors near $\tht=0$. 
From \eqref{expr:A_phi^pm} and \eqref{bkg_near0}, $\cD_q(\tht)$ defined in \eqref{def:cD_q} behaves as
\begin{align}
 \cD_q(\tht) &\simeq -\sRf r_0D_0\tht^2+r_0\zeta_{\rm f}, 
\end{align}
where 
\begin{align}
 \zeta_{\rm f} &\equiv \frac{1}{r_0}\brkt{q+\frac{1}{2}+\sRf\tl{\xi}}, 
\end{align}
and thus the mode equations in \eqref{md_eq:tlh} become 
\begin{align}
 \sbk{\der_\tht^2+\frac{\zeta_{\rm f}}{\tht^2}+\sRf D_0+\tl{\lmd}_q^2-\frac{\brkt{\zeta_{\rm f}-\sRf D_0\tht^2}^2}{\tht^2}}\tl{h}_{\rm R}(\tht;q) 
 &\simeq 0, \nonumber\\
 \sbk{\der_\tht^2-\frac{\zeta_{\rm f}}{\tht^2}-\sRf D_0+\tl{\lmd}_q^2-\frac{\brkt{\zeta_{\rm f}-\sRf D_0\tht^2}^2}{\tht^2}}\tl{h}_{\rm L}(\tht;q+1) 
 &\simeq 0, 
 \label{ap:md_eq:tlh}
\end{align}
where $\tl{\lmd}_q\equiv b_0\lmd_q/n_0$. 
When $q\neq -\sRf\tl{\xi}-1/2$, these are further approximated as
\begin{align}
 \sbk{\der_\tht^2+\frac{\zeta_{\rm f}\brkt{1-\zeta_{\rm f}}}{\tht^2}}\tl{h}_{\rm R}(\tht;q) &\simeq 0, \nonumber\\
 \sbk{\der_\tht^2-\frac{\zeta_{\rm f}\brkt{1+\zeta_{\rm f}}}{\tht^2}}\tl{h}_{\rm L}(\tht;q+1) &\simeq 0, 
\end{align}
whose solutions are 
\begin{align}
 \tl{h}_{\rm R}(\tht;q) &\simeq c_1\tht^{\zeta_{\rm f}}+c_2\tht^{1-\zeta_{\rm f}}, \nonumber\\
 \tl{h}_{\rm L}(\tht;q+1) &\simeq \tl{c}_1\tht^{-\zeta_{\rm f}}+\tl{c}_2\tht^{1+\zeta_{\rm f}}, 
\end{align}
where $c_1$, $c_2$, $\tl{c}_1$ and $\tl{c}_2$ are integration constants. 
When $q=-\sRf \tl{\xi}-1/2$, \eqref{ap:md_eq:tlh} becomes
\begin{align}
 \sbk{\der_\tht^2+\sRf D_0+\tl{\lmd}_q^2}\tl{h}_{\rm R}(\tht;q) &\simeq 0, \nonumber\\
 \sbk{\der_\tht^2-\sRf D_0+\tl{\lmd}_q^2}\tl{h}_{\rm L}(\tht;q+1) &\simeq 0, 
\end{align}
which are solved as
\begin{align}
 \tl{h}_{\rm R}(\tht;q) &\simeq c_1\cos\brkt{\sqrt{\tl{\lmd}_q^2+\sRf D_0}\tht}
 +c_2\sin\brkt{\sqrt{\tl{\lmd}_q^2+\sRf D_0}\tht}, \nonumber\\
 \tl{h}_{\rm L}(\tht;q+1) &\simeq \tl{c}_1\cos\brkt{\sqrt{\tl{\lmd}_q^2-\sRf D_0}\tht}
 +\tl{c}_2\sin\brkt{\sqrt{\tl{\lmd}_q^2-\sRf D_0}\tht}. 
\end{align}

Considering the finiteness of the mode functions at $\tht=0$, we obtain the following behaviors. 
\begin{align}
 \tl{h}_{\rm R}(\tht;q) &= \begin{cases} \cO\brkt{\tht^{\zeta_{\rm f}}} & \mbox{$0\leq \zeta_{\rm f}\leq \frac{1}{2}$ or $\zeta_{\rm f}>1$} \\
 \cO\brkt{\tht^{1-\zeta_{\rm f}}} & \mbox{$\zeta_{\rm f}<0$ or $\frac{1}{2}<\zeta_{\rm f}\leq 1$} \end{cases}, \nonumber\\
 \tl{h}_{\rm L}(\tht;q+1) &= \begin{cases} \cO\brkt{\tht^{-\zeta_{\rm f}}} & \mbox{$\zeta_{\rm f}<-1$ or $-\frac{1}{2}<\zeta_{\rm f}\leq 0$} \\
 \cO\brkt{\tht^{1+\zeta_{\rm f}}} & \mbox{$-1\leq \zeta_{\rm f}\leq -\frac{1}{2}$ or $\zeta_{\rm f}>0$} \end{cases}. 
\end{align}

Next we consider the behaviors near $\tht=\pi$. 
Since $\cD_q(\tht)$ behaves as
\begin{align}
 \cD_q(\tht) &\simeq \sRf r_\pi D_\pi\brkt{\pi-\tht}^2+r_\pi\alp_{\rm f}, 
\end{align}
where
\begin{align}
 \alp_{\rm f} &\equiv \frac{1}{r_\pi}\brkt{q+\frac{1}{2}-k_{\rm f}-\sRf \tl{\xi}}, 
\end{align}
the mode equations in \eqref{md_eq:tlh} are approximated as
\begin{align}
 &\sbk{\der_\tht^2-\frac{\alp_{\rm f}}{\brkt{\pi-\tht}^2}+\sRf D_\pi+\hat{\lmd}_q^2
 -\frac{\brc{\sRf D_\pi\brkt{\pi-\tht}^2+\alp_{\rm f}}^2}{\brkt{\pi-\tht}^2}}
 \tl{h}_{\rm R}(\tht;q) \simeq 0, \nonumber\\
 &\sbk{\der_\tht^2+\frac{\alp_{\rm f}}{\brkt{\pi-\tht}^2}-\sRf D_\pi+\hat{\lmd}_q^2
 -\frac{\brc{\sRf D_\pi\brkt{\pi-\tht}^2+\alp_{\rm f}}^2}{\brkt{\pi-\tht}^2}}
 \tl{h}_{\rm L}(\tht;q+1) \simeq 0, 
 \label{ap2:md_eq:tlh}
\end{align}
where $\hat{\lmd}_q=b_\pi\lmd_q/n_\pi$. 
When $q\neq k_{\rm f}+\sRf \tl{\xi}-1/2$, these are further approximated as
\begin{align}
 \sbk{\der_\tht^2-\frac{\alp_{\rm f}\brkt{1+\alp_{\rm f}}}{\brkt{\pi-\tht}^2}}\tl{h}_{\rm R}(\tht;q) &\simeq 0, \nonumber\\
 \sbk{\der_\tht^2+\frac{\alp_{\rm f}\brkt{1-\alp_{\rm f}}}{\brkt{\pi-\tht}^2}}\tl{h}_{\rm L}(\tht;q+1) &\simeq 0, 
\end{align}
and are solved as
\begin{align}
 \tl{h}_{\rm R}(\tht;q) &\simeq c_1\brkt{\pi-\tht}^{-\alp_{\rm f}}+c_2\brkt{\pi-\tht}^{1+\alp_{\rm f}}, \nonumber\\
 \tl{h}_{\rm L}(\tht;q+1) &\simeq \tl{c}_1\brkt{\pi-\tht}^{\alp_{\rm f}}+\tl{c}_2\brkt{\pi-\tht}^{1-\alp_{\rm f}}, 
\end{align}
with the integration constants~$c_1$, $c_2$, $\tl{c}_1$ and $\tl{c}_2$. 
When $q=k_{\rm f}+\sRf \tl{\xi}-1/2$, \eqref{ap2:md_eq:tlh} becomes
\begin{align}
 \sbk{\der_\tht^2+\sRf D_\pi+\hat{\lmd}_q^2}\tl{h}_{\rm R}(\tht;q) &\simeq 0, \nonumber\\
 \sbk{\der_\tht^2-\sRf D_\pi+\hat{\lmd}_q^2}\tl{h}_{\rm L}(\tht;q+1) &\simeq 0, 
\end{align}
whose solutions are 
\begin{align}
 \tl{h}_{\rm R}(\tht;q) &\simeq c_1\cos\brkt{\sqrt{\hat{\lmd}_q^2+\sRf D_\pi}\brkt{\pi-\tht}}
 +c_2\sin\brkt{\sqrt{\hat{\lmd}_q^2+\sRf D_\pi}\brkt{\pi-\tht}}, \nonumber\\
 \tl{h}_{\rm L}(\tht;q+1) &\simeq \tl{c}_1\cos\brkt{\sqrt{\hat{\lmd}_q^2-\sRf D_\pi}\brkt{\pi-\tht}}
 +\tl{c}_2\sin\brkt{\sqrt{\hat{\lmd}_q^2-\sRf D_\pi}\brkt{\pi-\tht}}. 
\end{align}
As a result, we obtain the following behaviors around $\tht=\pi$. 
\begin{align}
 \tl{h}_{\rm R}(\tht;q) &= \begin{cases} 
 \cO\brkt{\brkt{\pi-\tht}^{1+\alp_{\rm f}}} & \alp_{\rm f} < -\frac{1}{2} \\
 \cO\brkt{\brkt{\pi-\tht}^{-\alp_{\rm f}}} & \alp_{\rm f} \geq -\frac{1}{2} \end{cases}, \nonumber\\
 \tl{h}_{\rm L}(\tht;q+1) &= \begin{cases}
 \cO\brkt{\brkt{\pi-\tht}^{\alp_{\rm f}}} & \alp_{\rm f} < \frac{1}{2} \\
 \cO\brkt{\brkt{\pi-\tht}^{1-\alp_{\rm f}}} & \alp_{\rm f} \geq \frac{1}{2} 
 \end{cases}. 
\end{align}

\bibliography{Refs}{}

\providecommand{\href}[2]{#2}\begingroup\raggedright\begin{thebibliography}{10}

\bibitem{Arkani-Hamed:1998jmv}
N.~Arkani-Hamed, S.~Dimopoulos and G.R.~Dvali, \emph{{The Hierarchy problem and new dimensions at a millimeter}}, \href{https://doi.org/10.1016/S0370-2693(98)00466-3}{\emph{Phys. Lett. B} {\bfseries 429} (1998) 263} [\href{https://arxiv.org/abs/hep-ph/9803315}{{\ttfamily hep-ph/9803315}}].

\bibitem{Aghababaie:2003wz}
Y.~Aghababaie, C.P.~Burgess, S.L.~Parameswaran and F.~Quevedo, \emph{{Towards a naturally small cosmological constant from branes in 6-D supergravity}}, \href{https://doi.org/10.1016/j.nuclphysb.2003.12.015}{\emph{Nucl. Phys. B} {\bfseries 680} (2004) 389} [\href{https://arxiv.org/abs/hep-th/0304256}{{\ttfamily hep-th/0304256}}].

\bibitem{Lust:2019zwm}
D.~L\"ust, E.~Palti and C.~Vafa, \emph{{AdS and the Swampland}}, \href{https://doi.org/10.1016/j.physletb.2019.134867}{\emph{Phys. Lett. B} {\bfseries 797} (2019) 134867} [\href{https://arxiv.org/abs/1906.05225}{{\ttfamily 1906.05225}}].

\bibitem{Montero:2022prj}
M.~Montero, C.~Vafa and I.~Valenzuela, \emph{{The dark dimension and the Swampland}}, \href{https://doi.org/10.1007/JHEP02(2023)022}{\emph{JHEP} {\bfseries 02} (2023) 022} [\href{https://arxiv.org/abs/2205.12293}{{\ttfamily 2205.12293}}].

\bibitem{Otsuka:2022rpx}
H.~Otsuka and Y.~Sakamura, \emph{{Spacetime evolution during moduli stabilization in radiation dominated era beyond 4D effective theory}}, \href{https://doi.org/10.1007/JHEP08(2022)120}{\emph{JHEP} {\bfseries 08} (2022) 120} [\href{https://arxiv.org/abs/2205.00175}{{\ttfamily 2205.00175}}].

\bibitem{Otsuka:2022vgf}
H.~Otsuka and Y.~Sakamura, \emph{{Full higher-dimensional analysis of moduli oscillation and radiation in expanding universe}}, \href{https://doi.org/10.1007/JHEP05(2023)231}{\emph{JHEP} {\bfseries 05} (2023) 231} [\href{https://arxiv.org/abs/2212.14314}{{\ttfamily 2212.14314}}].

\bibitem{Otsuka:2024xsp}
H.~Otsuka and Y.~Sakamura, \emph{{Induced moduli oscillation by radiation and space expansion in a higher-dimensional model}}, \href{https://doi.org/10.1103/PhysRevD.109.115019}{\emph{Phys. Rev. D} {\bfseries 109} (2024) 115019} [\href{https://arxiv.org/abs/2402.15547}{{\ttfamily 2402.15547}}].

\bibitem{Salam:1984cj}
A.~Salam and E.~Sezgin, \emph{{Chiral Compactification on Minkowski x S**2 of N=2 Einstein-Maxwell Supergravity in Six-Dimensions}}, \href{https://doi.org/10.1016/0370-2693(84)90589-6}{\emph{Phys. Lett. B} {\bfseries 147} (1984) 47}.

\bibitem{Nishino:1984gk}
H.~Nishino and E.~Sezgin, \emph{{Matter and Gauge Couplings of N=2 Supergravity in Six-Dimensions}}, \href{https://doi.org/10.1016/0370-2693(84)91800-8}{\emph{Phys. Lett. B} {\bfseries 144} (1984) 187}.

\bibitem{Randjbar-Daemi:1985tdc}
S.~Randjbar-Daemi, A.~Salam, E.~Sezgin and J.A.~Strathdee, \emph{{An Anomaly Free Model in Six-Dimensions}}, \href{https://doi.org/10.1016/0370-2693(85)91653-3}{\emph{Phys. Lett. B} {\bfseries 151} (1985) 351}.

\bibitem{Green:1984bx}
M.B.~Green, J.H.~Schwarz and P.C.~West, \emph{{Anomaly Free Chiral Theories in Six-Dimensions}}, \href{https://doi.org/10.1016/0550-3213(85)90222-6}{\emph{Nucl. Phys. B} {\bfseries 254} (1985) 327}.

\bibitem{Kumar:2010ru}
V.~Kumar, D.R.~Morrison and W.~Taylor, \emph{{Global aspects of the space of 6D N = 1 supergravities}}, \href{https://doi.org/10.1007/JHEP11(2010)118}{\emph{JHEP} {\bfseries 11} (2010) 118} [\href{https://arxiv.org/abs/1008.1062}{{\ttfamily 1008.1062}}].

\bibitem{Aghababaie:2002be}
Y.~Aghababaie, C.P.~Burgess, S.L.~Parameswaran and F.~Quevedo, \emph{{SUSY breaking and moduli stabilization from fluxes in gauged 6-D supergravity}}, \href{https://doi.org/10.1088/1126-6708/2003/03/032}{\emph{JHEP} {\bfseries 03} (2003) 032} [\href{https://arxiv.org/abs/hep-th/0212091}{{\ttfamily hep-th/0212091}}].

\bibitem{Burgess:2011mt}
C.P.~Burgess and L.~van Nierop, \emph{{Large Dimensions and Small Curvatures from Supersymmetric Brane Back-reaction}}, \href{https://doi.org/10.1007/JHEP04(2011)078}{\emph{JHEP} {\bfseries 04} (2011) 078} [\href{https://arxiv.org/abs/1101.0152}{{\ttfamily 1101.0152}}].

\bibitem{Carroll:2003db}
S.M.~Carroll and M.M.~Guica, \emph{{Sidestepping the cosmological constant with football shaped extra dimensions}},  \href{https://arxiv.org/abs/hep-th/0302067}{{\ttfamily hep-th/0302067}}.

\bibitem{Burgess:2004ib}
C.P.~Burgess, \emph{{Towards a natural theory of dark energy: Supersymmetric large extra dimensions}}, \href{https://doi.org/10.1063/1.1848343}{\emph{AIP Conf. Proc.} {\bfseries 743} (2004) 417} [\href{https://arxiv.org/abs/hep-th/0411140}{{\ttfamily hep-th/0411140}}].

\bibitem{Williams:2012au}
M.~Williams, C.P.~Burgess, L.~van Nierop and A.~Salvio, \emph{{Running with Rugby Balls: Bulk Renormalization of Codimension-2 Branes}}, \href{https://doi.org/10.1007/JHEP01(2013)102}{\emph{JHEP} {\bfseries 01} (2013) 102} [\href{https://arxiv.org/abs/1210.3753}{{\ttfamily 1210.3753}}].

\bibitem{Vinet:2004bk}
J.~Vinet and J.M.~Cline, \emph{{Can codimension-two branes solve the cosmological constant problem?}}, \href{https://doi.org/10.1103/PhysRevD.70.083514}{\emph{Phys. Rev. D} {\bfseries 70} (2004) 083514} [\href{https://arxiv.org/abs/hep-th/0406141}{{\ttfamily hep-th/0406141}}].

\bibitem{Goldberger:1999uk}
W.D.~Goldberger and M.B.~Wise, \emph{{Modulus stabilization with bulk fields}}, \href{https://doi.org/10.1103/PhysRevLett.83.4922}{\emph{Phys. Rev. Lett.} {\bfseries 83} (1999) 4922} [\href{https://arxiv.org/abs/hep-ph/9907447}{{\ttfamily hep-ph/9907447}}].

\bibitem{Burgess:2012pc}
C.P.~Burgess, L.~van Nierop, S.~Parameswaran, A.~Salvio and M.~Williams, \emph{{Accidental SUSY: Enhanced Bulk Supersymmetry from Brane Back-reaction}}, \href{https://doi.org/10.1007/JHEP02(2013)120}{\emph{JHEP} {\bfseries 02} (2013) 120} [\href{https://arxiv.org/abs/1210.5405}{{\ttfamily 1210.5405}}].

\bibitem{Garriga:2000jb}
J.~Garriga, O.~Pujolas and T.~Tanaka, \emph{{Radion effective potential in the brane world}}, \href{https://doi.org/10.1016/S0550-3213(01)00144-4}{\emph{Nucl. Phys. B} {\bfseries 605} (2001) 192} [\href{https://arxiv.org/abs/hep-th/0004109}{{\ttfamily hep-th/0004109}}].

\bibitem{Lee:2007vy}
H.M.~Lee and A.~Papazoglou, \emph{{Supersymmetric codimension-two branes in six-dimensional gauged supergravity}}, \href{https://doi.org/10.1088/1126-6708/2008/01/008}{\emph{JHEP} {\bfseries 01} (2008) 008} [\href{https://arxiv.org/abs/0710.4319}{{\ttfamily 0710.4319}}].

\bibitem{Parameswaran:2006db}
S.L.~Parameswaran, S.~Randjbar-Daemi and A.~Salvio, \emph{{Gauge Fields, Fermions and Mass Gaps in 6D Brane Worlds}}, \href{https://doi.org/10.1016/j.nuclphysb.2006.12.020}{\emph{Nucl. Phys. B} {\bfseries 767} (2007) 54} [\href{https://arxiv.org/abs/hep-th/0608074}{{\ttfamily hep-th/0608074}}].

\end{thebibliography}\endgroup
\bibliographystyle{JHEP} 


\end{document}